\documentclass[10pt,english,twoside]{article}
\usepackage{authblk}

\pdfoutput=1
\usepackage{graphicx}
\usepackage{epstopdf, epsfig}

\usepackage{xcolor}
\usepackage{relsize}
\def\solidthick{\protect\rule[2pt]{10.pt}{1pt}}

\def\solidshort{\protect\rule[2pt]{3.pt}{1pt}}
\def\dashed{\solidshort$\,$\solidshort$\,$\solidshort}

\newcommand{\opencircle}{$\mathlarger{\mathlarger{\mathlarger{\circ}}}$}

\newcommand{\opensquare}{$\mathsmaller{\square}$}
\newcommand{\solidcircle}{$\mathlarger{\mathlarger{\mathlarger{\bullet}}}$}
\newcommand{\solidtriup}{$\blacktriangle$}

\newcommand{\solidtriright}{$\blacktriangleright$}
\newcommand{\solidtrileft}{$\blacktriangleleft$}

\newcommand{\solidsquare}{$\mathsmaller{\blacksquare}$}

\newcommand{\solidstar}{$\mathlarger{\mathlarger{\star}}$}

\definecolor{mygray}{rgb}{0.5,0.5,0.5}
\definecolor{col_110}{rgb}{0.1,0.1,0.9}
\definecolor{col_023}{rgb}{0.1,0.8,0.9}
\definecolor{col_120}{rgb}{0.1,0.3,0.2}
\definecolor{col_022}{rgb}{0.1,0.9,0.1}
\definecolor{col_100}{rgb}{0.9,0.1,0.1}
\definecolor{col_101}{rgb}{0.4,0.0,0.6}
\definecolor{col_034}{rgb}{1.0,0.5,0.0}
\definecolor{col_035}{rgb}{1.0,0.5,0.0}
\definecolor{col_036}{rgb}{1.0,0.5,0.0}
\definecolor{col_037}{rgb}{1.0,1.0,1.0}
\definecolor{col_gray}{rgb}{0.5,0.5,0.5}
\definecolor{my_dark_blue}{rgb}{0,0,0.5}

\definecolor{colmode_1_0}{rgb}{0.9,0.55,0.0}
\definecolor{colmode_2_0}{rgb}{1.0,0.0,0.0}
\definecolor{colmode_0_1}{rgb}{0.0,0.4,0.8}
\definecolor{colmode_1_1}{rgb}{1.0,0.0,1.0}
\definecolor{colmode_0_2}{rgb}{0.0,1.0,1.0}
\definecolor{colmode_1_2}{rgb}{0.6,0.42,0.8}
\definecolor{colmode_0_3}{rgb}{0.05,0.45,0.37}
\definecolor{colmode_1_3}{rgb}{0.0,0.0,0.0}
\definecolor{colmode_0_4}{rgb}{0.0,1.0,0.0}

\usepackage{amsmath,amssymb}

\usepackage{fancyhdr}

\usepackage[colorlinks,linkcolor=blue,citecolor=blue,urlcolor=blue]{hyperref}

\usepackage[round,sort&compress,semicolon]{natbib}
\usepackage{geometry}
\newlength\halflineskip
\newlength\affilskip
\usepackage{titlesec}
\titleformat{\section}{\large\bfseries}{\thesection}{1em}{}
\titleformat{\subsection}{\normalsize\bfseries}{\thesubsection}{1em}{}

\title{On the scaling of the instability of a flat sediment bed with respect to ripple-like patterns}
\author[1]{Markus Scherer}
\author[2]{Aman G. Kidanemariam}
\author[1]{Markus Uhlmann \footnote{Email address for correspondence: \href{markus.uhlmann@kit.edu}{markus.uhlmann@kit.edu}}}
\affil[1]{\small Institute for Hydromechanics, Karlsruhe Institute of Technology, 76131 Karlsruhe, Germany}
\affil[2]{\small Department of Mechanical Engineering, The University of Melbourne, Victoria 3010, Australia}
\vspace{-1.3cm}
\date{\small (Dated: \today \ -- original submission, revised version has been accepted \\ for publication in \textit{J.\ Fluid Mech.} 2020)}

\newcommand{\icaseA}{\ensuremath{H4D052}}
\newcommand{\icaseAA}{\ensuremath{H2D052}}
\newcommand{\icaseB}{\ensuremath{H6D052}}
\newcommand{\icaseC}{\ensuremath{H6D077}}
\newcommand{\icaseD}{\ensuremath{H6D102}}
\newcommand{\icaseE}{\ensuremath{H6D154}}

\newcommand{\icaseXc}{\ensuremath{H4D102^3}}
\newcommand{\icaseXabc}{\ensuremath{H4D102^{1,2,3}}}
\newcommand{\icaseY}{\ensuremath{H2D102^1}}
\newcommand{\icaseZ}{\ensuremath{H2D102^2}}
\begin{document}

\newcommand{\KU}{KU2017}

\maketitle
\thispagestyle{empty}
\vspace{-0.5cm}
%
\begin{abstract}
We investigate the formation of subaqueous transverse bedforms in turbulent open channel flow by means of direct numerical simulations with fully-resolved particles.
The main goal of the present analysis is to address the question whether the initial pattern wavelength scales with the particle diameter or with the mean fluid height.
A previous study (Kidanemariam \& Uhlmann, \textit{J. Fluid Mech.}, vol. 818, 2017, pp. 716-743) has observed a lower bound for the most unstable pattern wavelength in the range $75-100$ times the particle diameter, which was equivalent to $3-4$ times the mean fluid height.
In the current paper, we vary the streamwise box length in terms of the particle diameter and of the mean fluid height independently in order to distinguish between the two possible scaling relations.
For the chosen parameter range, the obtained results clearly exhibit a scaling of the initial pattern wavelength with the particle diameter, with a lower bound around a streamwise extent of approximately $80$ particle diameters. 
In longer domains, on the other hand, patterns are observed at initial wavelengths in the range $150-180$ times the particle diameter, which is in good agreement with experimental measurements.
Variations of the mean fluid height, on the other hand, seem to have no significant influence on the most unstable initial pattern wavelength.
Furthermore, for the cases with the largest relative submergence, we observe spanwise and streamwise sediment waves of similar amplitude to evolve and superimpose, leading to three-dimensional sediment patterns.

\end{abstract}

  \section{Introduction}\label{sec:intro}
Sediment beds under streaming water bodies often exhibit characteristic bedforms which can significantly affect the transport properties of the flow.
Among these various bedforms, transverse bedforms are usually classified as ripples, dunes or antidunes \citep{Yalin_1977}. Ripples and dunes both exhibit a roughly triangular shape with a gentle slope on the upstream surface and a steeper downstream face, the latter approximately inclined at the angle of repose \citep{Best_2005}. However, these two bedforms differ in their scaling properties: while ripples are small compared to the fluid height and thus their wavelength is believed to scale with the particle diameter, the height of dunes is large enough to distinctly modify the flow field over the entire flow depth. As a consequence, the amplitude of dunes should scale with the fluid height \citep{Engelund_Fredsoe_1982}.
In contrast to the remaining bedform types, antidunes exhibit a symmetric cross-section and they can, depending on the hydraulic conditions, travel in the upstream or the downstream direction. Antidunes can only form in free-surface flows, whereas ripples and dunes can form in configurations without a free surface as well, such as pipe flows \citep[see e.g.][]{Ouriemi_2009} or closed-conduit flows \citep[see e.g.][]{Coleman_al_2003,CardonaFlorez_Franklin_2016}.

The origin of these transverse bedforms has been found to be an instability process in which an initially flat sediment bed deforms after an initial perturbation in the flow system leading to the formation of transverse patterns \citep{Yalin_1977}. \citet{Inglis_1949} proposed that inhomogeneities of the sediment bed such as small particle agglomerations could act as initial perturbations that trigger subsequent bedform evolution.
In a more recent paper, \citet{Coleman_Melville_1996} proposed an instability process originating in an isolated random pileup that grows with time and, once it has reached a critical height of approximately $3-4$ times the particle diameter $D$, induces the generation of further pileups downstream.
\citet{Coleman_Nikora_2009,Coleman_Nikora_2011} describe the formation of the initial bedforms as a two-stage formation process. In a first stage, random sediment patches with a typical length of $7-15D$ interact on an active (but still planar) sediment bed due to sediment transport events that are believed to be related to coherent turbulent structures. Once the height of these initial random patches exceeds some threshold height, the initial disturbance stabilizes by agglomerating sediment particles, with the consequence that further downstream regular patterns of sand-wavelets form.
\citet{Venditti_al_2005} observed different modes of bedform initiation in their experiments depending on the flow rate and, consequently, on the Reynolds and Froude number. At lower flow rates, local defects are seen to initiate a similar formation cycle as that described by \citet{Coleman_Melville_1996}, whereas at higher flow rates, the authors observed patterns to spontaneously evolve over the entire bed, leading to a regular `cross-hatch pattern'.

Unfortunately, up to the present date, accurate measurements of the initial bedform dimensions remain challenging. On the one hand, this is due to the very short time window in which the initial bedform evolution can be observed, and, on the other hand, due to the small height of the initial patterns of only a few particle diameters which makes them hard to detect.
It is for this reason that most of the experimental studies in the last decades have focused on the description of fully-developped patterns in the equilibrium state \citep[see e.g.][]{Yalin_1985}, where the wavelength of the initial bedforms is typically much higher than in the initial phase \citep{Langlois_Valance_2007}. On the other hand, only a small number of experimentalists were able to quantitatively describe the wavelength of the initial bedforms.
In turbulent open channel flow, \citet{Coleman_Melville_1996} and \citet{Coleman_Nikora_2009} observed the initial wavelength to scale mainly with the sediment size and to be rather unaffected by the flow conditions. Similar observations were made by \citet{Coleman_al_2003} in turbulent closed conduit flows as well as by \citet{Coleman_Eling_2000} in laminar open channel flows, the latter indicating in particular that the formation of initial bedforms is not restricted to the turbulent regime and the accompanying turbulent bursts, as earlier suggested by \citet{Raudkivi_1997}.
Similarly, \citet{Langlois_Valance_2007} and \citet{CardonaFlorez_Franklin_2016} both report under turbulent channel flow conditions that the initial wavelength depends mainly on the particle diameter, while the influence of the shear velocity and the flow conditions is rather weak.
\citet{Ouriemi_2009}, on the contrary, measured initial wavelengths of the order of the fluid height in pipe flows, while \citet{Franklin_2008} observed a dependence of the initial wavelength on both the particle diameter and the shear velocity in experimental measurements of turbulent closed conduit flows.

In theoretical studies based on linear stability analysis, the flow system is usually described by a simplified model for the driving flow (such as RANS models or potential flow solutions) combined with a sediment bed continuity equation for the evolution of the bed \citep[see e.g.][]{Kennedy_1963,Kennedy_1969,Charru_2006}.
In order to close the system of equations, a formulation for the particle flux is required. In most stability analysis, it is assumed that the particle flux and the local bed shear stress are in phase, which allows then to express the particle flux as a function of the local bed shear stress \citep{Charru_al_2013}.
In recent studies, this assumption has been removed and additional relaxation equations are used to take into account a possible phase lag between both quantities \citep{Charru_2006}.
A linear stability analysis is then performed in order to determine regions of instability in the parameter space as well as the most amplified wavelength for a given flow configuration. During the past decades, a large number of stability analysis for different flow configurations including different stabilizing and destabilizing effects has been presented for turbulent \citep[see e.g.][]{Richards_1980,Sumer_Bakioglu_1984,Colombini_2004,Colombini_Stocchino_2011,Fourriere_al_2010} as well as for laminar flows \citep[see e.g.][]{Charru_Mouilleron-Arnould_2002,Charru_Hinch_2006}.
A detailed overview of the different approaches used in linear stability analysis can be found in the reviews of \citet{Engelund_Fredsoe_1982}, \citet{Seminara_2010} and \citet{Charru_al_2013}.
Recently, \citet{Zgheib_Balachandar_2019} have presented a combined numerical-theoretical stability analysis, in which the bed shear stress for the linear stability analysis is computed by means of direct numerical simulations (DNS) in which the sediment bed is represented in a continuous and impermeable fashion.

However, despite an increasing complexity of the used models, the wavelengths predicted by most linear stability analysis still differs by more than one order of magnitude from the wavelengths observed in experiments \citep{Langlois_Valance_2007,Ouriemi_2009}.
Furthermore, the observations concerning the scaling of the initial wavelengths differ markedly between the different studies.
For instance, \citet{Fourriere_al_2010} found a single region with unstable wavelengths independent of the fluid height, leading them to the conclusion that only ripples can form directly from a flat bed, while dunes develop by a ripple coarsening processes only. \citet{Colombini_Stocchino_2011}, in contrast, observed two separate regions of instability with most amplified wavelengths of the size of the fluid height and of the particle diameter, respectively, which suggests that either ripples or dunes may form out of the same instability mechanism.

It should be kept in mind that the validity of linear stability investigations is limited to the very first instances of sediment bed evolution \citep{Coleman_Melville_1996}, i.e. the theory is not able to predict the bedform dimensions correctly, once non-linear effects become dominant. In more recent studies, thus, weakly non-linear analysis have been presented \citep[e.g.][]{Colombini_Stocchino_2008} to take into account also non-linear effects.

In recent years an alternative method for the investigation of sediment transport in its early stages has become available in form of DNS featuring fully-resolved particles, which allow to resolve all relevant flow scales even below the particle length scale \citep[e.g.][]{Kidan_Uhlmann_2014a,Kidan_Uhlmann_2014b,Kidan_Uhlmann_2017,Derksen_2015,
Vowinckel_2014,Vowinckel_2017b}.
In a set of numerical experiments, \citet{Kidan_Uhlmann_2017} (henceforth denoted \KU{}) have recently determined a lower bound for the minimal unstable wavelength, $\lambda_{th}$, by reducing the streamwise domain length $L_x$ successively in a comparable concept as the minimal flow unit of \citet{Jimenez_Moin_1991}. It was observed that below a threshold $L_x = \lambda_{th}$, the formation of transverse bedforms is effectively hindered and a perturbed bed remains stable as a consequence of the limited domain size, even though the conditions would otherwise allow the bed to become unstable.
For the considered parameter values, \KU{} found $\lambda_{th}/D$ to be in the range $75-100$, which is equivalent to a range of $\lambda_{th}/H_f = 3-4$ in their cases. Since, however, the relative fluid height $H_f/D$ has been kept constant over all simulations, it was not possible to further distinguish between the two alternative scaling relations.

The purpose of the present work is to investigate the scaling of the lower threshold for the minimal unstable wavelength.
In order to be able to distinguish between the two alternative scalings, i.e. either with the particle diameter $D$ or the fluid height $H_f$, two new series of numerical experiments are performed in which the relative streamwise domain lengths $L_x/D$ and $L_x/H_f$ are varied independently.
In the subsequent analysis, we investigate the influence of both length scales on the stability of the subaqueous bedforms using the newly computed DNS data. The present manuscript is organized as follows.
In \S~\ref{sec:numerics}, we briefly describe the numerical method which we use for the simulation of subaqueous sediment transport in this work. An overview over the relevant physical and numerical parameters as well as the chosen flow configurations is given in \S~\ref{sec:flow_config}.
Subsequently, in \S~\ref{sec:interf_extr}, we shortly present two different ways of defining the fluid-bed interface, depending on whether the sediment bed is analysed in a spanwise-averaged framework or in its full extent.
In \S~\ref{sec:results}, we analyse the data obtained in the context of the two simulation series separately, focussing on the bedform geometry and its temporal evolution.
A discussion of the results as well as a comparison with experimental and theoretical studies follows in \S~\ref{sec:discussion}.
We conclude the study with a summary of the main findings in \S~\ref{sec:conclusion}.

  \section{Numerical method}\label{sec:numerics}
%
%
%
We use the same numerical method used in \citet{Kidan_Uhlmann_2014a,Kidan_Uhlmann_2014b,Kidan_Uhlmann_2017} to solve the coupled fluid-solid problem. For the simulation of the fluid phase, the incompressible Navier-Stokes equations are solved numerically in the entire computational domain using a second order finite difference scheme together with a fractional step algorithm on a uniform Cartesian grid. Time integration of the governing equations is done in a semi-implicit way, including a Crank-Nicholson scheme for the viscous terms and a low-storage three-step Runge-Kutta scheme for the non-linear terms.
The immersed boundary formulation of \citet{Uhlmann_2005} is then used to couple the flow field with the solid phase: localized force terms are introduced into the Navier-Stokes equations which impose the no-slip condition at the interface between fluid and solid phase.
The motion of the particles is obtained by time integration of the Newton-Euler equations for rigid-body motion. The driving force and torque comprises hydrodynamic and gravitational contributions as well as those resulting from particle-particle and particle-wall contact.
Due to the fact that the characteristic time scale of particle collisions is typically several orders of magnitude smaller than those of the turbulent fluid motion, a sub-stepping method is used for the numerical time integration of the Newton-Euler equations \citep{Kidan_Uhlmann_2014b}.

Momentum exchange due to particle collisions is computed using a soft-sphere discrete-element model (DEM). In the frame of the chosen approach, two particles are defined as `being in contact', if the minimal distance between their surfaces $\Delta$ falls below a force range $\Delta_c$. In this case, a contact force and torque act on both particles which is defined as the sum of three individual contributions, i.e. an elastic normal force, a normal damping force and a tangential frictional force. The elastic normal force is a linear function of the penetration length $\delta_c = \Delta_c -\Delta$ with a constant stiffness coefficient $k_n$. The normal damping force is a linear function of the normal component of the relative particle velocity between the two particles at the contact point, with a constant normal damping coefficient $c_n$. Similarly, the tangential frictional force is defined as a linear function of the tangential component of the relative particle velocity at the contact point, with a constant tangential friction coefficient $c_t$. Note that the magnitude of the tangential frictional force has an upper traction limit in the form of the Coulomb friction limit with a friction coefficient $\mu_c$. A detailed description of the collision model and extensive validation can be found in \citet{Kidan_Uhlmann_2014b}.

For each simulation, the model thus requires the choice of the four force parameters ($k_n$,$c_n$,$c_t$,$\mu_c$) as well as the force range $\Delta_c$. Introducing a dry restitution coefficient $\varepsilon_d$, which is defined as the absolute value of the ratio between the normal components of the relative velocity before and after a dry collision, allows to relate the normal stiffness coefficient $k_n$ and the normal damping coefficient $c_n$, and to formulate the model depending on the alternative set of force parameters ($k_n$,$\varepsilon_d$,$c_t$,$\mu_c$).
Due to the varying particle size and submerged weight in the present simulations, the set of parameters used in \citet{Kidan_Uhlmann_2014a} and \citet{Kidan_Uhlmann_2017} has been adapted to the respective cases. The force range $\Delta_c$ is set equal to the uniform grid spacing $\Delta x$ for all cases with a particle diameter $D \leq 15 \Delta x$ and equal to $2\Delta x$ for all cases with larger particles. The stiffness parameter $k_n$ has a value between approximately $8400$ and $17000$ times the submerged weight of a single particle, divided by the particle diameter in the respective case.
The parameters are chosen such that the maximum overlap $\delta_c$ is within a few percent of $\Delta_c$. The dry restitution coefficient is set to $\varepsilon_d=0.3$, which determines, together with the chosen value of $k_n$, the constant normal damping coefficient $c_n$. The constant tangential friction coefficient is set to the same value $c_t = c_n$. Finally, the Coulomb friction coefficient was fixed at $\mu_c = 0.4$ except for cases \icaseAA{} and \icaseD{}, in which a slightly higher value $\mu_c = 0.5$ was unintentionally used.
In Appendix \ref{sec:append_A} we show, however, that this difference in the limiting Coulomb friction has only a minor influence on the eventually developed bedform and that it does not affect the stability or instability of the sediment bed.

  \section{Flow configuration and parameter values}\label{sec:flow_config}
\begin{figure}
      \includegraphics[width=\textwidth]{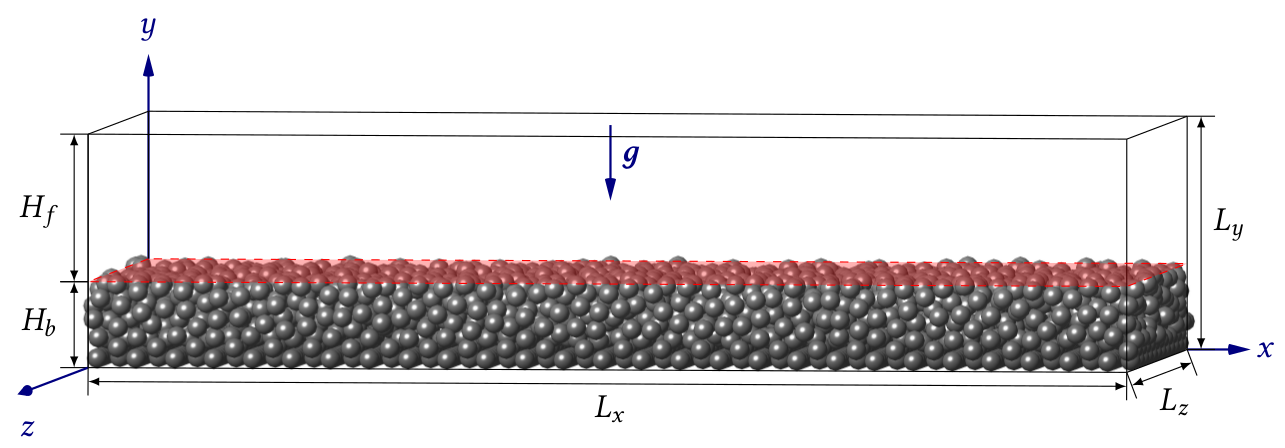}
	  \caption{Schematic of the open channel flow configuration.
            Flow is in the positive $x$-direction.
            The computational domain is periodic along the $x$- and
            $z$-directions. No-slip and free-slip boundary conditions are
            imposed at the bottom ($y=0$) and top ($y=L_y$), respectively.
            }
   \label{fig:flowconfig_sketch}
\end{figure}

\begin{table}
  \begin{center}
%

%
%
\resizebox{\textwidth}{!}{%
 \begin{tabular}{l c c c c c c c c l}
  \multicolumn{1}{c}{Case}&
  \multicolumn{1}{c}{$Re_b$}&
  \multicolumn{1}{c}{$Re_\tau$}&
  \multicolumn{1}{c}{$\rho_p/\rho_f$}&
  \multicolumn{1}{c}{$Ga$}&
  \multicolumn{1}{c}{$D^+$}&
  \multicolumn{1}{c}{$H_f/D$}&
  \multicolumn{1}{c}{$H_b/D$}&
  \multicolumn{1}{c}{$\theta$}&
  \multicolumn{1}{l}{source}\\
  \\
 {{\color{black}   $\solidthick$}}\icaseAA{}& 3011 & 242.2 & 2.5 &  28.37 &  9.59 & 25.25 & 13.15 & 0.11 & \textit{present}\\[1.5pt]
 {{\color{black}   $\solidthick$}}\icaseA{} & 3011 & 281.0 & 2.5 &  56.74 & 22.67 & 12.40 &  6.80 & 0.16 & \textit{present}\\[1.5pt]
 \\
 {{\color{col_110} $\solidthick$}}\icaseB{} & 3011 & 311.1 & 2.5 & 104.24 & 39.62 &  7.86 &  4.94 & 0.14 & \textit{present}\\[1.5pt]
 {{\color{col_023} $\solidthick$}}\icaseC{} & 3011 & 286.1 & 2.5 &  56.74 & 23.18 & 12.34 &  6.86 & 0.17 & \textit{present}\\[1.5pt]
 {{\color{col_120} $\solidthick$}}\icaseD{} & 3011 & 279.4 & 2.5 &  43.61 & 17.10 & 16.33 &  9.27 & 0.15 & \textit{present}\\[1.5pt]
 {{\color{col_022} $\solidthick$}}\icaseE{} & 3011 & 303.0 & 2.5 &  28.37 & 11.90 & 25.47 & 12.93 & 0.18 & $H6$ in \KU\\
 \\[1.5pt]
 {{\color{col_036} $\solidthick$}}\icaseXabc{} & 3011 & 267.3 & 2.5 &  28.37 & 10.63 & 25.15 & 13.25 & 0.14 & $H4^{1,2,3}$ in \KU\\
 {{\color{col_100} $\solidthick$}}\icaseY{} & 5013 & 360.5 & 2.5 &  16.78 &  7.14 & 50.53 & 13.47 & 0.18 & \textit{present}\\[1.5pt]
 {{\color{col_101} $\solidthick$}}\icaseZ{} & 5013 & 358.5 & 2.5 &  20.06 &  7.09 & 50.59 & 13.41 & 0.13 & \textit{present}\\[1.5pt]
\end{tabular}}
  \caption{Physical parameters of the simulations. $Re_b$, $Re_{\tau}$ and $D^+$ are the bulk Reynolds number, the friction Reynolds number and the particle Reynolds number, respectively. The density ratio $\rho_p/\rho_f$ and the Galileo number $Ga$ are imposed in each simulation, whereas the Shields number $\theta$, the relative submergence $H_f/D$ and the relative sediment bed height $H_b/D$ are computed a posteriori (cf. table~\ref{tab:param_numer}).
  The last column provides information about the source of the listed cases, dinstinguishing between simulations that have been computed in the course of the current study (\textit{present}) and cases that are from \citet{Kidan_Uhlmann_2017}(\KU{}).
  It should be further mentioned that the physical parameters presented for case \icaseXabc{} have been averaged over the three individual simulations. A list of the physical parameters for each individual run can be found in \KU{}.
  }
\label{tab:param_phys}


\hrulefill \\
\hfill \\

%
%
\resizebox{\textwidth}{!}{%
   \begin{tabular}{l r c c c r c c}
     \multicolumn{1}{c}{Case}&
     \multicolumn{1}{c}{$[L_x \times L_y \times L_z]/D$}&
     \multicolumn{1}{c}{$L_x/H_f$}&
     \multicolumn{1}{c}{$D/\Delta x$}&
     \multicolumn{1}{c}{$\Delta x^+$}&
     \multicolumn{1}{c}{$N_p$}&
     \multicolumn{1}{c}{$T_{obs}/T_b$}&
     \multicolumn{1}{c}{$T^s_{obs}/T_b$}\\
     \\
     {{\color{black}   $\solidthick$}}\icaseAA{}& $ 51.2 \times 38.4 \times 76.8$ & 2.03 & 10 & 0.96 &  43 730 & 742 & 691\\[1.5pt]
     {{\color{black}   $\solidthick$}}\icaseA{} & $ 51.2 \times 19.2 \times 38.4$ & 4.13 & 20 & 1.13 &   9 923 & 497 & 177\\[1.5pt]
     \\
     {{\color{col_110} $\solidthick$}}\icaseB{} & $ 51.2 \times 12.8 \times 25.6$ & 6.52 & 30 & 1.32 &   5 086 & 879 & 405\\[1.5pt]
     {{\color{col_023} $\solidthick$}}\icaseC{} & $ 76.8 \times 19.2 \times 38.4$ & 6.22 & 20 & 1.16 &  14 954 & 918 & 486\\[1.5pt]
     {{\color{col_120} $\solidthick$}}\icaseD{} & $102.4 \times 25.6 \times 51.2$ & 6.27 & 15 & 1.14 &  44 163 & 618 & 235\\[1.5pt]
     {{\color{col_022} $\solidthick$}}\icaseE{} & $153.6 \times 38.4 \times 76.8$ & 6.04 & 10 & 1.19 & 127 070 & 918 & 513\\[1.5pt]
     \\
     {{\color{col_036} $\solidthick$}}\icaseXabc{} & $102.4 \times 38.4 \times 76.8$ & 4.07 & 10 & 1.06 &  86 645 & 401/400/853 & 88/84/542\\[1.5pt]
     {{\color{col_100} $\solidthick$}}\icaseY{} & $102.4 \times 64.0 \times 76.8$ & 2.03 & 10 & 0.71 &  86 645 & 736 & 393\\[1.5pt]
     {{\color{col_101} $\solidthick$}}\icaseZ{} & $102.4 \times 64.0 \times 76.8$ & 2.02 & 10 & 0.71 &  86 645 & 509 & 252\\[1.5pt]
   \end{tabular}}
   \caption{Numerical parameters of the simulations. The computational domain has dimensions $L_i$ in the $i-$th direction and is discretized using a uniform grid with mesh width $\Delta x$, $\Delta x^+$ denoting the grid width in wall-units. $N_p$ is the total number of particles in the respective case.
   The time is scaled in bulk time units $T_b=H_f/u_b$.
   $T_{obs}$ is the total observation time of each simulation starting from the release of the moving particles.
   Time dependent physical and numerical parameters in tables~\ref{tab:param_phys} and \ref{tab:param_numer} ($Re_{\tau}$, $D^+$, $H_f$, $H_b$, $\theta$, $\Delta x^+$) are computed as an average over a final time interval $T^s_{obs}$.
   }
\label{tab:param_numer}
  \end{center}
\end{table}

%
%
%
In the course of the current study, we have performed seven new independent direct numerical simulations of bedform
evolution over an erodible subaqueous sediment bed in a turbulent open channel. Two additional cases from \KU{} have been included in our analysis. The studied flow configuration is shown in figure~\ref{fig:flowconfig_sketch}. As can be seen, a Cartesian coordinate system is centered at the lower boundary of the open channel such that the $x$-, $y$- and $z$-direction are the streamwise, wall-normal and spanwise direction, respectively. Mean flow is directed in positive $x$-direction and gravity in the negative $y$-direction. In the streamwise and spanwise directions, periodic boundary conditions are imposed, whereas a no-slip- and a free slip-condition are imposed at the bottom and at the top plane of the channel, respectively.

As characteristic length scales, we define the mean fluid height $H_f$ and the mean sediment bed height $H_b$ as an average over both spatial directions (streamwise and spanwise) as well as over time, i.e. $H_f=\langle h_f \rangle_{zxt}$ and $H_b=\langle h_b \rangle_{zxt}$, respectively.
A precise definition of $h_f(x,z,t)$ and $h_b(x,z,t)$ will be given in section \ref{subsec:1D_interface}. Before the simulations with mobile particles are started, sediment bed and flow field have completed a start-up procedure, which is described in detail in \citet{Kidan_Uhlmann_2014a}.
The names of the individual simulations are chosen according to their streamwise domain length in terms of the mean fluid height and in terms of the particle diameter. For instance, in case \icaseA{}, the relative streamwise box length is $L_x/H_f \approx 4$ and $L_x/D = 51.2$, respectively (cf. table~\ref{tab:param_phys} and table~\ref{tab:param_numer}).

In all cases, the flow is driven by a time-dependent streamwise pressure gradient, that is adjusted at each time step to ensure a constant flow rate $q_f$. Therefore, the bulk Reynolds number can be computed \textit{a priori} as ${Re_b = u_b H_f/\nu}$, where the bulk velocity is defined as $u_b \equiv q_f/H_f$.
The friction Reynolds number is defined as ${Re_{\tau} = u_{\tau} H_f/\nu}$, where the friction velocity $u_{\tau}$ is computed \textit{a posteriori} from the streamwise and spanwise averaged total mean shear stress, which is composed of viscous stresses, turbulent Reynolds stresses and the stresses resulting from the fluid-particle interaction.
Due to the absence of a horizontal bottom wall, we evaluate the mean friction velocity $u_{\tau}$ at the location of the mean fluid-bed interface $y=H_b$, which can be interpreted as a virtual wall. For a more detailed description of the determination of the shear stress distribution, the reader is referred to \KU{}.

From dimensional analysis, it can be concluded that by adding sediment to a turbulent flow, the parameter space increases and in total, four non-dimensional numbers are required to fully describe a given system. In addition to $Re_b$, we choose the density ratio $\rho_p/\rho_f$, the non-dimensional length scale $H_f/D$ as well as the Galileo number ${Ga = u_{g} D/\nu}$, which expresses the ratio between gravity and viscous forces, where the gravitational velocity scale is $u_g= \sqrt{\left(\rho_{p}/\rho_{f}-1\right)|\mathbf{g}| D}$.

In the present simulations, we set the density ratio at $\rho_p/\rho_f = 2.5$ which is comparable to the values reported for glass in water. To allow pattern formation, the non-dimensional boundary shear stress, expressed by the Shields number ${\theta = u_{\tau}^{2}/u_{g}^{2}= ( D^{+} /Ga)^2}$, has to be larger than the critical value for incipient sediment motion. In turbulent flows, the critical Shields number has been observed to be in a range $\theta_c = 0.03-0.05$, slightly depending on the Galileo number \citep{Soulsby_Whitehouse_1997,Wong_Parker_2006,Franklin_Charru_2011}.

The non-dimensional mean fluid height $H_f/D$ is varied in the different simulations to elucidate the relevant length scales that dominate the scaling of subaqueous bedforms by either increasing the particle diameter or the mean fluid height while keeping the same dimensions for the domain. As a consequence, the number of fully-resolved particles lies in a range between $\mathcal{O}(10^3)$ in the largest particle case and up to $\mathcal{O}(10^5)$ in the case with the smallest particles.

Note that in all cases, the dimensions of the computational domain are sufficiently large to allow self-sustained turbulence. In particular, the case with the shortest streamwise and spanwise dimensions (scaled in viscous lengths) is case \icaseZ{} with $L_x^+ \approx 726$ and $L_z^+ \approx 555$. By comparison, \citet{Jimenez_Moin_1991} report the dimensions of their minimal flow-unit as
$L_x^+ \approx 250-350$ and $L_x^+ \approx 100$, respectively.

In section \ref{sec:results}, we will analyse the two simulation series in which either the particle diameter or the mean fluid height have been varied separately. It should be noted that we have performed two additional simulations \icaseAA{} and \icaseA{}, which do not fit into either of these two series. Therefore, we discuss the results obtained in these latter cases in section \ref{sec:discussion} only.

  \section{Extraction of bedform dimensions}\label{sec:interf_extr}
  For the following definition of the fluid-bed interface, we will consider the domain as composed of two distinct regions, that are, a lower particle-dominated region, hereafter termed as \textit{the sediment bed}, and the overlaying fluid-dominated region. Hence, the wall-normal dimension of the channel height $L_y$ can be written as the sum of the instantaneous local height of the fluid phase $h_f(x,z,t)$ and that of the sediment bed $h_b(x,z,t)$, \textit{viz.}
\begin{equation}\label{eqn:channel_height_decomp}
L_y = h_b(x,z,t) + h_f(x,z,t).
\end{equation}
In the chosen Cartesian coordinate system, the wall-normal location of the fluid-bed interface is identical to the value of the function $h_b(x,z,t)$. In the following, we will present two different approaches to define the instantaneous fluid-bed interface. The first method defines the fluid-bed interface through a spanwise averaging, whereas the second extracts the sediment bed as a two-dimensional surface.

%
%
%
%
%
%
\subsection{Definition and analysis of the spanwise-averaged fluid-bed interface} \label{subsec:1D_interface}
Assuming statistical homogeneity of the observed system, equation~\eqref{eqn:channel_height_decomp} can be averaged in the spanwise direction as
\begin{equation}\label{eqn:channel_height_decomp_zav}
L_y = \langle h_b \rangle_{z} (x,t) + \langle h_f \rangle_{z} (x,t).
\end{equation}
The spanwise averaged sediment bed height $\langle h_b \rangle_{z}$ is determined depending on a threshold for the solid volume fraction. Since the method is presented in detail in our previous works \citep{Kidan_Uhlmann_2014a,Kidan_Uhlmann_2017}, we restrict ourself to a short summary of the most important points. First, a solid-phase indicator function $\phi_p(\mathbf{x},t)$ is defined, which attains the value of unity for Eulerian grid points being located inside the particle domain $\Omega_p$ and zero elsewhere.
Second, the spanwise-averaged sediment bed height $\langle h_b \rangle_{z}$ is defined as the wall-normal location, at which the spanwise averaged solid indicator function $\langle \phi_p \rangle_{z}$ attains a threshold of $\langle \phi_p \rangle_{z}^{thresh}=0.1$ \citep{Kidan_Uhlmann_2014b}, i.e.
\begin{equation}
\langle h_b \rangle_{z} (x,t) = y \quad | \, \langle \phi_p \rangle_{z} (x,y,t)=\langle \phi_p \rangle_{z}^{thresh}.
\end{equation}
The spanwise-averaged fluid height $\langle h_f \rangle_{z}$ is computed using relation~\eqref{eqn:channel_height_decomp_zav}.
Eventually, we perform another decomposition in the streamwise direction, similar to the spanwise decomposition used in equation \eqref{eqn:channel_height_decomp_zav}. The resulting expression
\begin{equation}\label{eqn:01_hb_decomp}
\langle h_b \rangle_{z} (x,t) = \langle h_b \rangle_{zx} (t) + \langle h_b \rangle_{z}' (x,t),
\end{equation}
divides the spanwise-averaged interface in an instantaneous mean height $\langle h_b \rangle_{zx} (t)$ and a fluctuation $\langle h_b \rangle_{z}' (x,t)$ with respect to the former. In the following, the fluctuation will form the basis for the analysis of the bedform evolution.
The size and shape of two-dimensional transverse patterns is usually quantified by the pattern length and height. Over the last decades, a variety of different approaches has been developed to define these length scales. An overview over some of these methods is presented in \citet{Coleman_Nikora_2011}. In the current work, we choose a statistical definition for the pattern height as well as for the wavelength of the bedforms \citep{Langlois_Valance_2007}. As a measure for the pattern height, we use the root mean square of the sediment bed height fluctuation
\begin{equation}\label{eqn:rms_1D}
\sigma_{h}(t)=\sqrt{ \left\langle h_{b}^{\prime}(x, t) \cdot h_{b}^{\prime}(x, t)\right\rangle_{x} }.
\end{equation}
In order to determine the average pattern wavelength, let us define the instantaneous two-point correlation coefficient of the sediment bed height fluctuation
\begin{equation}\label{eqn:2pointcorr_1D}
R_{h}(\delta x,t)=\left\langle h_{b}^{\prime}(x, t) \cdot h_{b}^{\prime}(x+\delta x, t)\right\rangle_{x}/\sigma_{h}^{2}(t),
\end{equation}
where $\delta x$ expresses the streamwise separation length between two points.
With increasing $\delta x$, the evolution of $R_{h}(\delta x,t)$ at a given time $t$ behaves similar to a damped oscillation curve around a zero mean. The global (and first) minimum of the curve occurs at a separation $\delta x_{min}$, i.e.
\begin{equation}\label{eqn:wlength_1D}
R_{h}(\delta x_{min},t) \leq R_{h}(\delta x,t) \quad \forall \; \delta x \;  \in \; [0,Lx/2].
\end{equation}
The mean wavelength is finally obtained as twice this separation length, i.e. ${\lambda_h(t) = 2 \delta x_{min}}$.
%
%
%
%
%
%
\subsection{Definition and analysis of the two-dimensional fluid-bed interface} \label{subsec:2D_interface}
The spanwise-averaged definition and analysis, as described in the previous section, is restricted to a fluid-bed interface that is statistically homogeneous in the spanwise direction. However, in the general case, this procedure may be not adequate to describe possible three-dimensional sediment patterning.

Here, we use a criterion based on determining the top-layer particles of the fluid-bed interface. In classical morphodynamics, the sediment bed is distinguished from bedload and suspended load \citep{Bagnold_1956,VanRijn_1984}. Following this classification, the developed algorithm sorts out the latter two categories, leaving only the particles that are part of the sediment bed itself. First, single suspended particles are detected based on the wall-normal collision force component $F_{c,y}$. While particles inside the bed are exposed to a permanent wall-normal contact force of the order of their submerged weight $F_w=(\rho_p-\rho_f) \, \pi D^3/6 \, |\mathbf{g}|$, single suspended particles are typically not in contact with surrounding particles and, accordingly, only a small contact force acts on them. Thus, particles that are only exposed to a negligible wall-normal contact force ($|F_{c,y}|/F_w < 10^{-5}$) will not be considered as `bed particles'.

The remaining particles include both the actual sediment bed and the bedload layer. Particles inside the latter are in motion close to the bed, such that they lose the contact to the bed only for short time intervals \citep{Yalin_1977}. In order to separate the bedload transport particles from the bed particles, a criterion based upon the particle speed is employed, eliminating all particles which exceed a threshold value. The threshold value can be chosen by defining a non-dimensional particle Shields number $\theta_p = (|\mathbf{u_p}|/u_g)^2$ with the norm of the particle velocity $|\mathbf{u_p}|$ and the gravitational velocity scale $u_g$. The sediment bed particles are then found as the set of all particles for which $\theta_p \leq \theta_c$. Here we choose as the critical value $\theta_c=0.05$ \citep{Wong_Parker_2006}.
Note that the particle velocity criterion also eliminates suspended particle pairs which might instantaneously experience collision forces above the chosen threshold of the wall-normal contact force.

The fluid-bed interface is then defined through a three-dimensional $\alpha$-shape \citep{Edelsbrunner_Muecke_1994} enclosing the set of sediment bed particles. This enclosing surface can be thought of as a generalization of a convex hull, with an imposed radius $\alpha$ (here taken as $1.1$ times the particle diameter) that defines the length scale above which non-convexity is allowed.
The fluid-bed interface is then made up of those triangular faces of the $\alpha$-shape which have an outward-pointing face-normal with a positive $y$-component. The data consists of a function $h(x,z)$ which is sampled at the projection of the vertex points of the surface triangulation upon the $(x,z)$-plane. In a final step, this function $h(x,z)$ is interpolated to a uniform Eulerian grid with a mesh width equal to one particle diameter and the result is smoothed using a two-dimensional box-filter with a width of $5$ particle diameters.

For the analysis of this two-dimensional interface, let us extend our definition of the root mean square of the sediment bed height fluctuation, as defined in equation \eqref{eqn:rms_1D}, to the multidimensional case in the following form:
\begin{equation}\label{eqn:rms_2D}
\sigma_h^{2D} = \sqrt{\langle \, h_b' (x,z,t) \cdot h_b' (x,z,t) \, \rangle_{zx}}.
\end{equation}

  \section{Results}\label{sec:results}
    \subsection{Variation of the particle diameter}\label{sec:results_Dvar}
    In a first series, the influence of the particle diameter $D$ on the minimal unstable wavelength $\lambda_{th}$ at a given parameter point is investigated. To this end, a series of four simulations with different particle size is performed, in which the particle diameter $D$ is successively increased from $D/H_f \approx 0.04$ in case \icaseE{} to $D/H_f \approx 0.13$ in case \icaseB{}. It should be stressed that in all four simulations, the streamwise domain length $L_x$ is almost constant in the range $6.04 - 6.52 H_f$ and thus clearly above the critical value observed in \KU{}. Note that case \icaseE{} with the smallest particle diameter is identical to case $H6$ presented in \KU{}, in which a single bedform has been observed to evolve and to eventually reach a quasi-equilibrium state.

\begin{figure}[t]
    \centering
         \begin{minipage}{2ex}
           \rotatebox{90}
           {\small \hspace{6ex}$ z/D$}
         \end{minipage}
         \begin{minipage}{.47\linewidth}%
           \raggedright{(\textit{a})}
           \includegraphics[width=\linewidth]
           {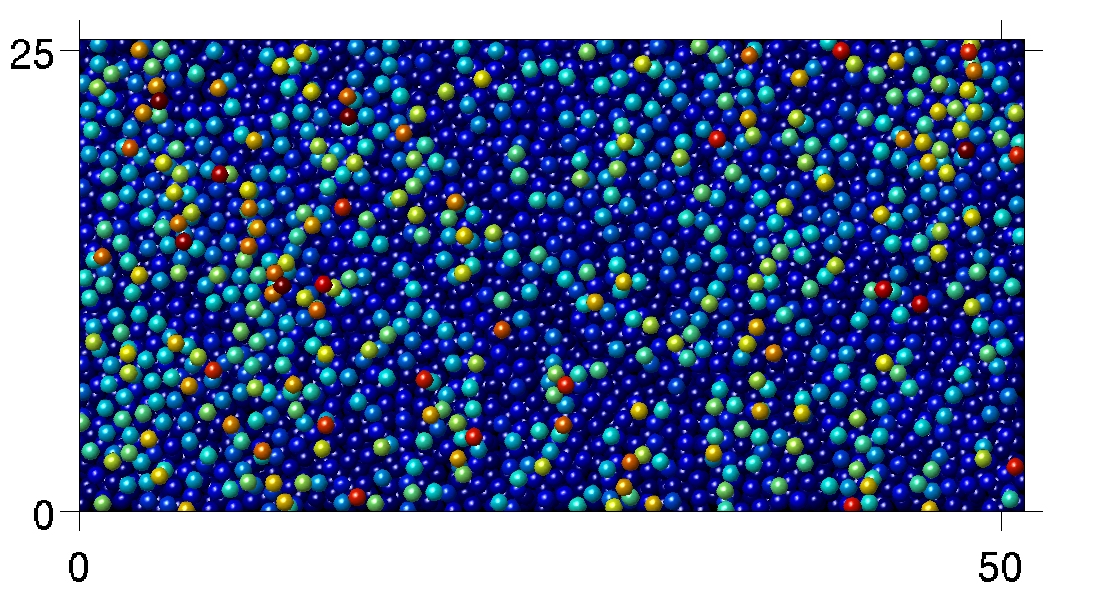}
         \end{minipage}
         \begin{minipage}{.47\linewidth}%
           \raggedright{(\textit{b})}
           \includegraphics[width=\linewidth]
           {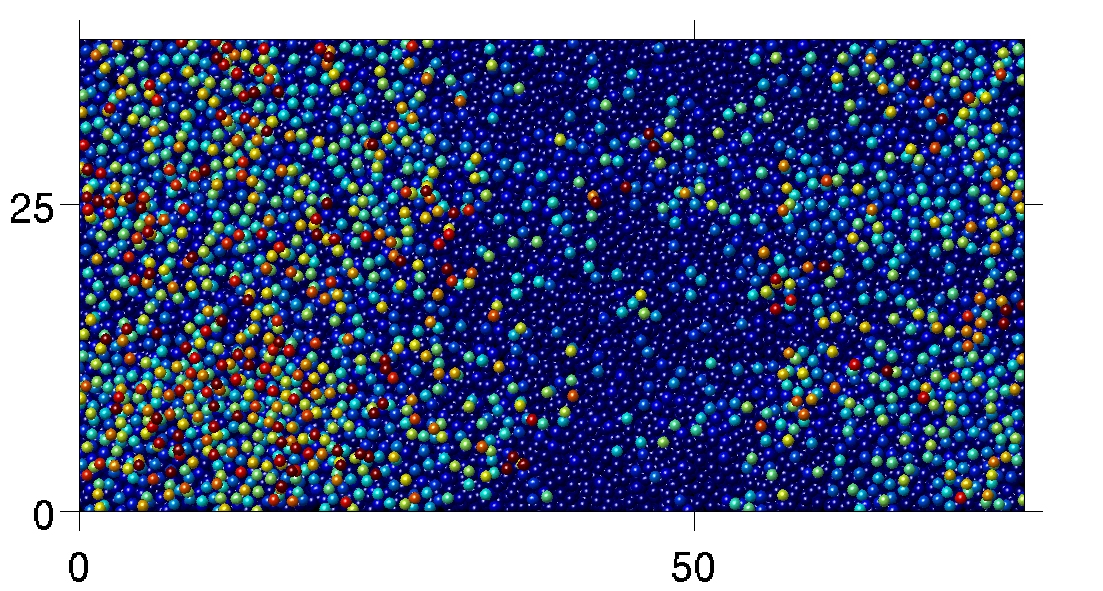}
         \end{minipage}
         \begin{minipage}{2ex}
           \rotatebox{90}
           {\small \hspace{6ex}$ z/D$}
         \end{minipage}
         \begin{minipage}{.47\linewidth}
           \raggedright{(\textit{c})}
           \includegraphics[width=\linewidth]
           {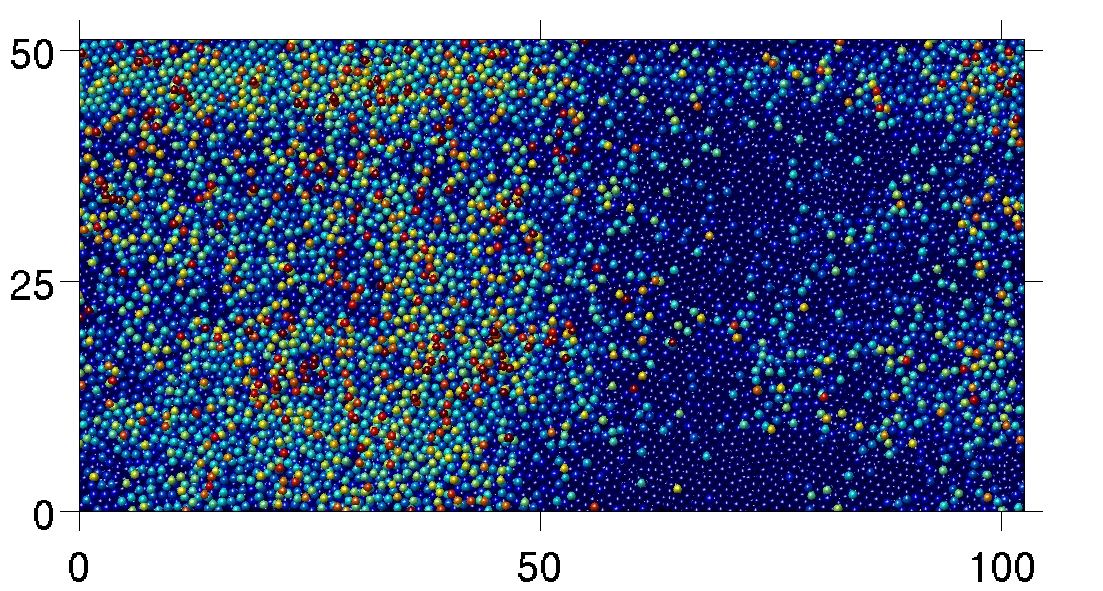}
           \centerline{\small $ x/D$}
         \end{minipage}
          \begin{minipage}{.47\linewidth}
           \raggedright{(\textit{d})}
           \includegraphics[width=\linewidth]
           {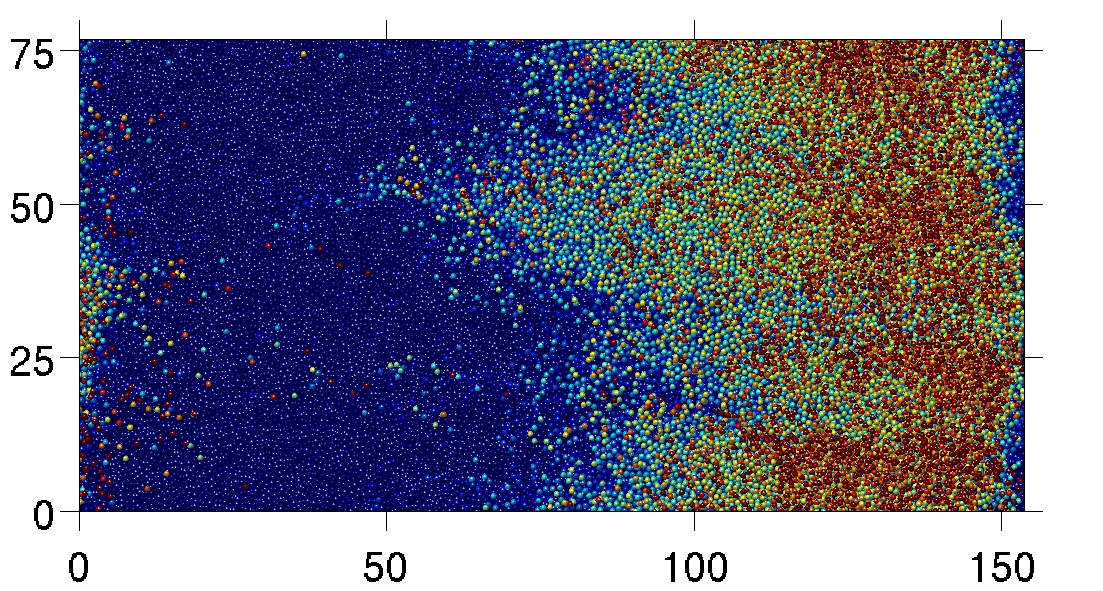}
           \centerline{\small $ x/D$}
         \end{minipage}
         \hfill
         { }\\[3ex]
         \begin{minipage}{0.33\linewidth}%
           \includegraphics[width=\linewidth]
           {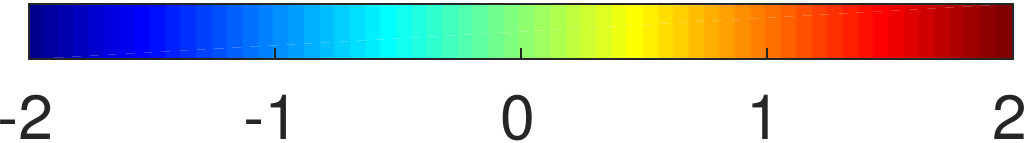}
           \centerline{\small $ (y-H_b)/D$}
         \end{minipage}
\caption{Instantaneous snapshots of the sediment bed of the cases varying the particle diameter, seen from the top of the channel. Particles are coloured depending on the wall-normal location of their centre, as shown in the global color code.
The bedforms shown in the figures have been observed at $t \approx 300 T_b$ for all cases:
(\textit{a}) \icaseB{},
(\textit{b}) \icaseC{},
(\textit{c}) \icaseD{},
(\textit{d}) \icaseE{}.
Supplementary movies are available at
\href{https://dx.doi.org/10.4121/uuid:7eb6a0be-ff83-4883-9d99-31daaa6a2863}{https://dx.doi.org/10.4121/uuid:7eb6a0be-ff83-4883-9d99-31daaa6a2863}.
}
\label{fig:part_snapshots_D}
\end{figure}

%
%
Figure \ref{fig:part_snapshots_D} shows instantaneous snapshots of the sediment bed in the different cases after a simulation period of approximately $300$ bulk time units.
It is seen that in case \icaseB{}, no transverse pattern has formed. Instead, the bed has remained essentially flat and eroded sediment seems to distribute over the entire channel length and width, without accumulating in specific locations. In particular, case \icaseB{} does not exhibit a similar regular pattern of streamwise aligned alternating ridges and troughs as those \KU{} found in the stable sediment bed of their case $H3$.
On the other hand, the remaining new cases with a domain length $L_x \geq 76.8D$ are unstable, each featuring one single transverse bedform with an initial wavelength of the order of the streamwise box length. This indicates that at the given parameter point, there is indeed a lower limit for $\lambda_{th}$, which depends on the particle diameter and which is found in the range $ 51.2 - 76.8D$.
As can already be inferred from the top-view visualizations of the sediment bed (cf. figure \ref{fig:part_snapshots_D}), the shape of those bedforms which do emerge differs, in particular in the vicinity of the threshold.

\begin{figure}[t]
    \centering
         \begin{minipage}{2ex}
           \rotatebox{90}
           {\small \hspace{5ex}$\sigma_h /D$}
         \end{minipage}
         \begin{minipage}{.46\linewidth}
           \raggedright{(\textit{a})}
           \includegraphics[width=\linewidth]
           {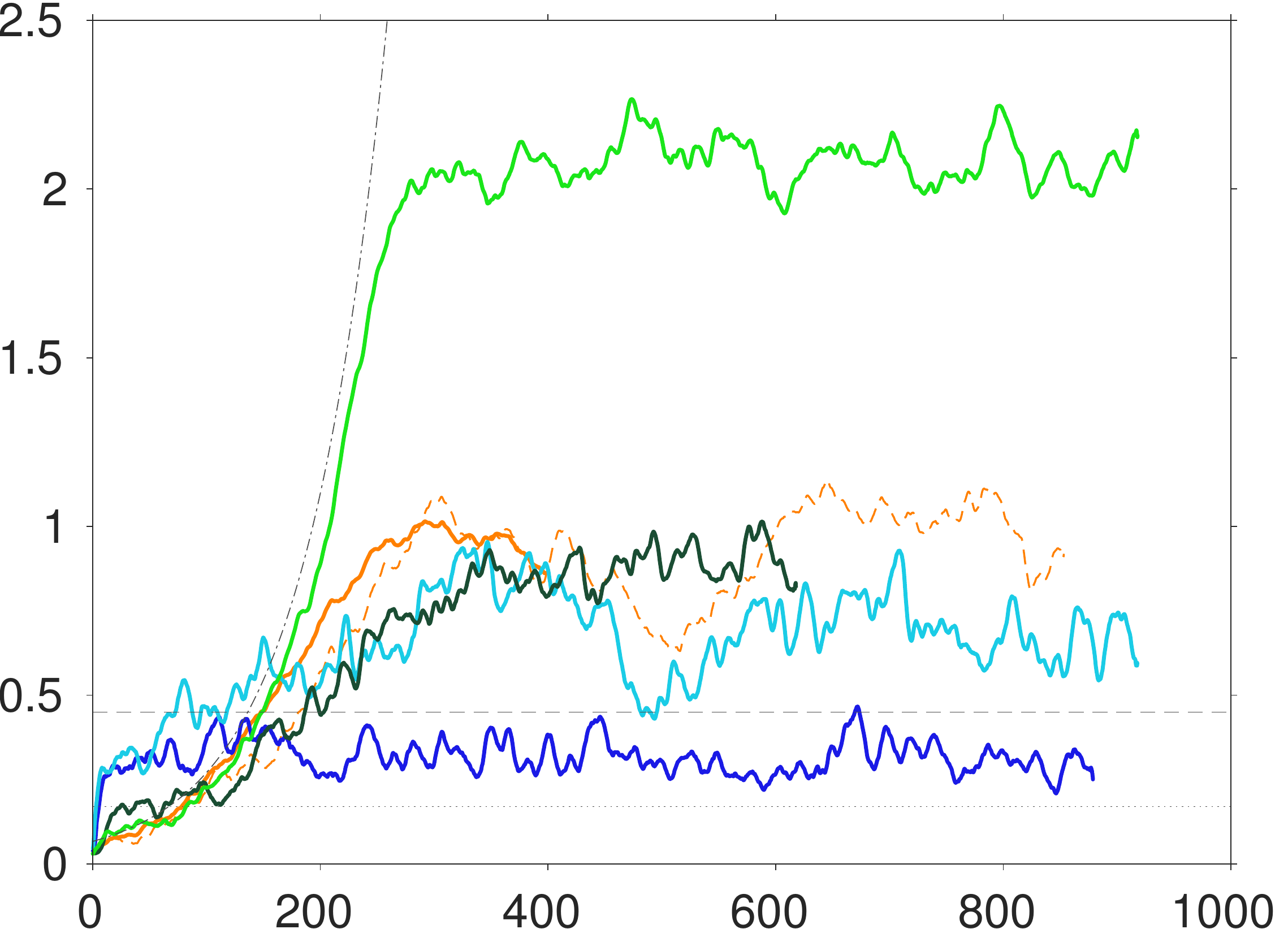}
           \centerline{\small $ t/(H_f/u_b)$}
         \end{minipage}
         \begin{minipage}{.46\linewidth}
           \raggedright{(\textit{b})}
           \includegraphics[width=\linewidth]
           {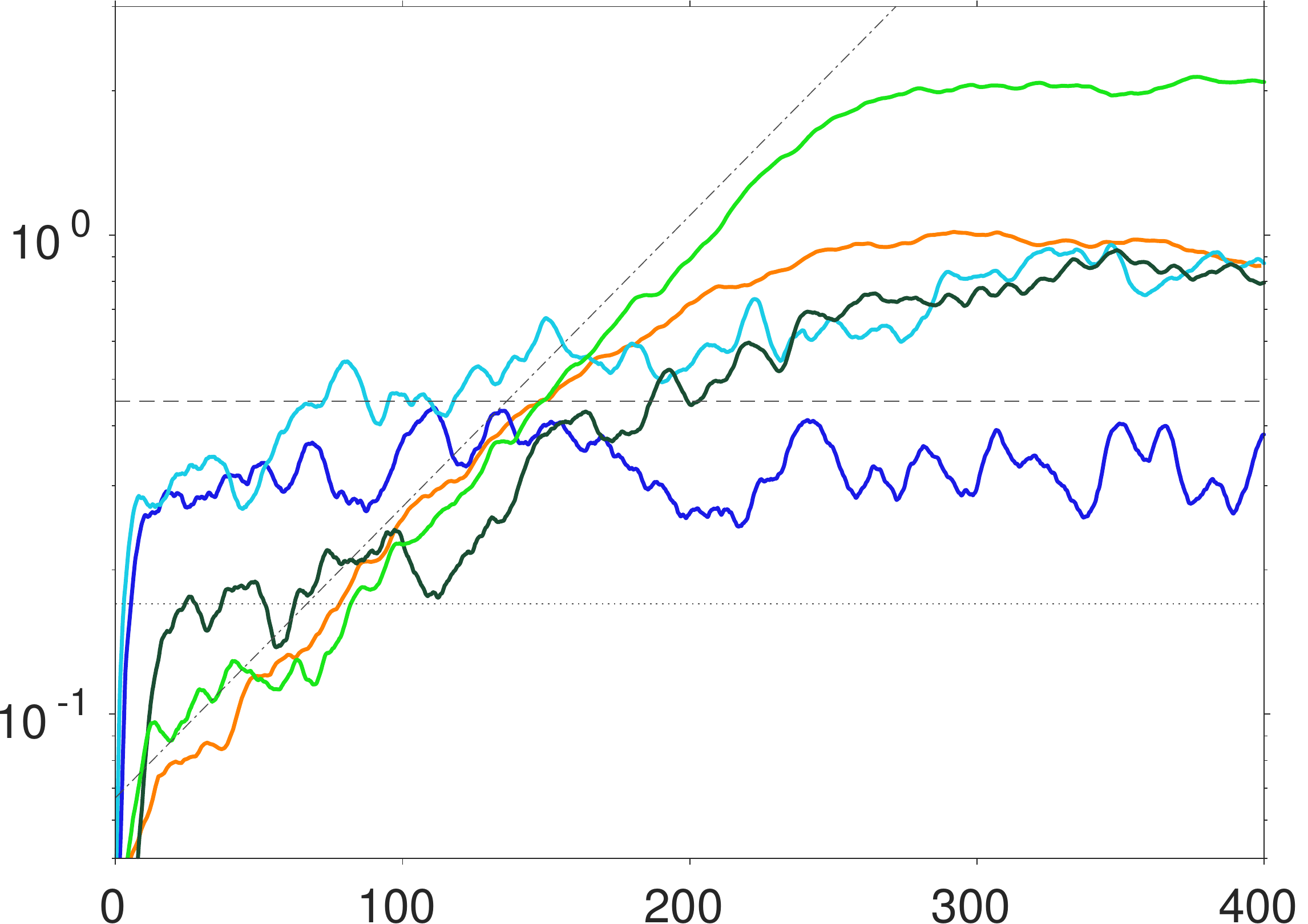}
           \centerline{\small $ t/(H_f/u_b)$}
         \end{minipage}
  	\caption{(\textit{a}) Time evolution of the root mean square of the bedform amplitude normalized by the particle diameter $\sigma_h /D$.
    Time is scaled in bulk time units $T_b$.
    Cases
    \icaseB{} ({{\color{col_110} $\solidthick$}}),
    \icaseC{} ({{\color{col_023} $\solidthick$}}),
    \icaseD{} ({{\color{col_120} $\solidthick$}}),
    \icaseE{} ({{\color{col_022} $\solidthick$}}).
    The data of case \icaseXabc{} is presented as ensemble average over the three simulations ({{\color{col_034} $\solidthick$}}, thick line). Additionaly, the individual evolution of run \icaseXc{} is presented ({{\color{col_034} $\dashed$}}, thin line).
    The horizontal dotted and dashed lines indicate values reported by \citet{Coleman_Nikora_2009} for
    a `static plane bed' ($\sigma_h \approx 0.17D$)
    and `mobile sediments on planar but active beds' ($\sigma_h \approx 0.40 - 0.50D$, here $\sigma_h \approx 0.47D$).
    The dashed-dotted line represents the exponential curve $\sigma_h/D = 0.0668 \exp(0.0140t/T_b)$ found by \KU{} as the best fit for the initial growth of $\sigma_h$ in their cases (including the present case \icaseE{}).
    Mean values of $\sigma_h$ averaged over the final time interval $T^s_{obs}$ are as follows:
    \icaseB{}: $\langle \sigma_h \rangle_t/D \approx 0.30$,
    \icaseC{}: $\langle \sigma_h \rangle_t/D \approx 0.67$,
    \icaseD{}: $\langle \sigma_h \rangle_t/D \approx 0.89$,
    \icaseE{}: $\langle \sigma_h \rangle_t/D \approx 2.08$,
    \icaseXabc{}: $\langle \sigma_h \rangle_t/D \approx 0.96$.
    (\textit{b}) Same data as (\textit{a}), but represented in semi-logarithmical scale.
    }
\label{fig:RMS1D_Dvar}
\end{figure}

In order to provide a quantitative description of these patterns, we will now discuss various geometrical measures. First, we will focus on the pattern height evolution by studying the root mean square sediment bed height fluctuation $\sigma_h$, which can be seen as a measure for the inhomogeneity of the sediment bed height \citep{Kidan_Uhlmann_2014a,Kidan_Uhlmann_2017,Zgheib_al_2018b}. In a case, in which no transverse bedforms evolve, $\sigma_h$ will show some fluctuations due to random uncorrelated bed undulations, but it will remain bounded by a small value in the course of the simulation. A sediment bed that shows such evolution will be classified as stable. In unstable systems, on the other hand, $\sigma_h$ will grow more or less monotonically during the initial phase of the simulation and in particular, it will exceed the aforementioned threshold that bounds the stable cases.
%
%
Figure \ref{fig:RMS1D_Dvar} shows the time evolution of $\sigma_h$.
After a short initial transient of a few bulk time units during which the chaotic particle motion leads to small finite values of $\sigma_h$, we observe that in all cases with $L_x \geq 76.8D$, $\sigma_h$ increases with time starting at $t \approx 20 T_b$, indicating that the chosen relative box length is sufficiently long to cover at least one unstable mode and thus to allow the bed to evolve transverse patterns.
On the contrary, $\sigma_h$ in case \icaseB{} does not exhibit any substantial growth period. Instead, it remains around a value of $\sigma_h \approx 0.3D$, only featuring fluctuations of small amplitude, bounded by an upper value of approximately $0.47D$. This value is in good agreement with the results of \citet{Coleman_Nikora_2009}, who observed ``mobile sediments on planar but active bed'' for $\sigma_h$ in the range $0.40 - 0.50D$.
Unstable bedforms are usually observed to run through different phases of bedform evolution (\KU{}). During an initial growth period, the bed increases exponentially as predicted by linear stability theory. After some time, non-linear contributions become relevant and let the bed height tend to its quasi-steady equilibrium value \citep{Charru_al_2013}.
\KU{} observed that all their unstable cases exhibited an exponential growth at a very similar growth rate, apparently independent of the streamwise box length of the respective case. They found an exponential function
\begin{equation}
\sigma_h/D = A \exp(B t/T_b),
\end{equation}
with amplitude $A=0.0668$ and growth rate $B=0.0140$ to best fit the initial increase of $\sigma_h$ with time over an interval of approximately $150$ bulk time units.
In contrast, the temporal evolution in the first instances of the current unstable cases strongly varies from case to case.
While the initial growth of case \icaseE{} follows the above described exponential function of \KU{}, the remaining two cases \icaseD{} and \icaseC{} grow more slowly.
In the subsequent phase, case \icaseE{} and \icaseD{} attain what could be called as asymptotic state after approximately $280$ and $350$ bulk time units, respectively. Case \icaseC{} exhibits stronger oscillations with higher amplitude compared to the previous two cases, a behaviour similar to that of case \icaseXc{} from \KU{}.
Averaging $\sigma_h$ over the final time interval $T^s_{obs}$ leads to values of $\langle \sigma_h \rangle_t/D \approx 2.08$, $\langle \sigma_h \rangle_t/D \approx 0.89$ and $\langle \sigma_h \rangle_t/D \approx 0.67$ for cases \icaseE{}, \icaseD{} and \icaseC{}, respectively.
It is remarkable that the mean values in the final interval differ by more than a factor of two between cases \icaseE{} and \icaseD{},
which possess a comparable relative domain length of $L_x/H_f \approx 6$.
On the other hand, cases \icaseD{} and \icaseXabc{} attain very similar values of $\langle \sigma_h \rangle_t/D \approx 0.89$ and $\langle \sigma_h \rangle_t/D \approx 0.96$, respectively, although case \icaseXabc{} has a smaller relative box length $L_x/H_f \approx 4$.
This indicates that the attained mean value of $\sigma_h$ in the final interval mainly depends on the chosen $L_x/D$ ratio, whereas it is not very sensitive to a variation of the mean fluid height $H_f$.

\begin{figure}[t]
    \centering
         \begin{minipage}{2ex}
           \rotatebox{90}
           {\small \hspace{5ex}$\lambda_h/D$}
         \end{minipage}
         \begin{minipage}{.46\linewidth}
           \includegraphics[width=\linewidth]
           {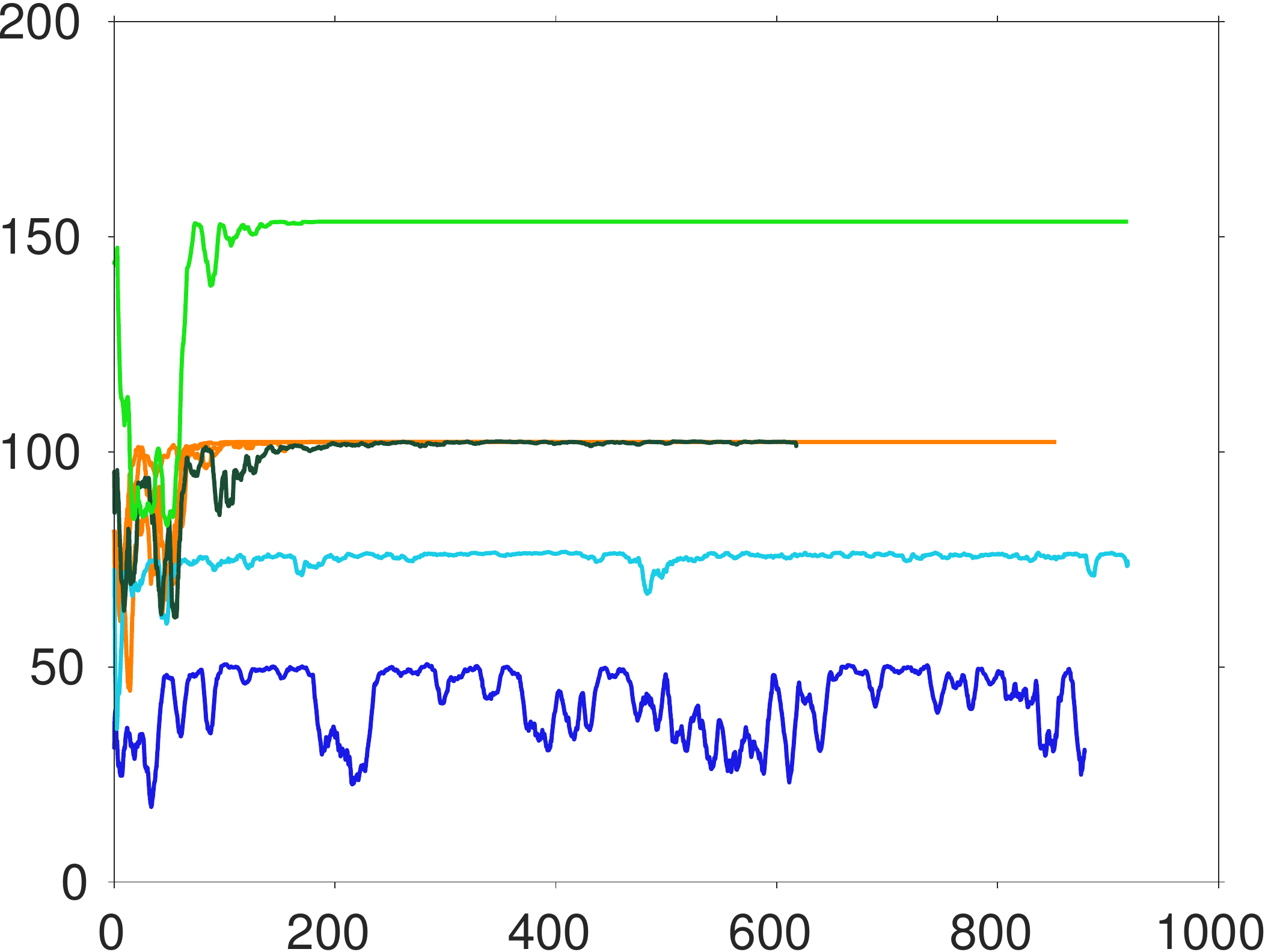}
           \centerline{\small $ t/(H_f/u_b)$}
         \end{minipage}
  	\caption{Time evolution of the mean wavelength of the sediment bed height normalized by the particle diameter $\lambda_h /D$.
    Colour coding similar to figure \ref{fig:RMS1D_Dvar}:
    Cases
     \icaseB{} ({{\color{col_110} $\solidthick$}}),
     \icaseC{} ({{\color{col_023} $\solidthick$}}),
     \icaseD{} ({{\color{col_120} $\solidthick$}}),
     \icaseE{} ({{\color{col_022} $\solidthick$}}),
     \icaseXabc{} ({{\color{col_034} $\solidthick$}}).
     }
\label{fig:wavelength1D_Dvar}
\end{figure}

%
%
Figure \ref{fig:wavelength1D_Dvar} shows the time evolution of the mean pattern wavelength $\lambda_h$.
It can be observed that the chosen mean wavelength for all cases with $L_x \geq 76.8D$ settle, after some fluctuation in the first $100 - 200$ bulk time units, at the maximum possible wavelength, i.e. $\lambda_h = L_x$, and maintains this value until the end of the simulation.
\KU{} observed a similar evolution of the mean wavelength settling at $\lambda_h = L_x$ for their shorter cases up to a box length $L_x = 179.2D$. The current observations further support their findings that these systems cannot freely choose their initial wavelength, but that they are constrained by the limitation of the streamwise domain size. As a consequence, the system chooses the maximum possible unstable wavelength as the dominant one, which, however, is not necessary the same as the one it would choose in a system without spatial limitations (cf. the very long box with $L_x/H_f=48$ simulated by \KU{}).
In contrast to the observed unstable cases, $\lambda_h$ in case \icaseB{} jumps between the first three harmonics $\lambda_1 = L_x$, $\lambda_2 = L_x/2$ and $\lambda_3 = L_x/3$ for the entire observation interval. This behaviour is caused by random uncorrelated bed disturbances and it indicates that the system is not able to develop a pattern at a finite wavelength.

%
%
%
In the following, we will investigate the fully-developed spanwise-averaged profile of the sediment pattern along the streamwise direction.
To this end, we compute the phase-averaged bed height $\tilde{H}_b(\tilde{x})$ as defined in \KU{}. For this purpose we use a constant pattern migration velocity $u_D$ which is determined from a linear fit of the space-time correlation.
Phase-averaging is performed over the final time interval $T^s_{obs}$ (as indicated in table~\ref{tab:param_numer}).
We further introduce the aspect ratio $AR$ and the degree of asymmetry $LR$ of the pattern as (\KU{})

\begin{subequations}
\begin{align}
AR &= H_D/\lambda_h \\
LR &= l_D/\lambda_h,
\end{align}
\end{subequations}
where $H_D=max(\tilde{H}_b)-min(\tilde{H}_b)$ is the pattern height, $l_D$ is the streamwise distance from the crest to the neighbouring downstream trough and $\lambda_h$ is the mean wavelength.

\begin{figure}[t]
    \centering
         \begin{minipage}{2ex}
           \rotatebox{90}
           {\small \hspace{4ex}$(\tilde{H}_b - min(\tilde{H}_b))/\lambda_h$}
         \end{minipage}
         \begin{minipage}{.45\linewidth}
           \raggedright{(\textit{a})}
           \includegraphics[width=\linewidth]
           {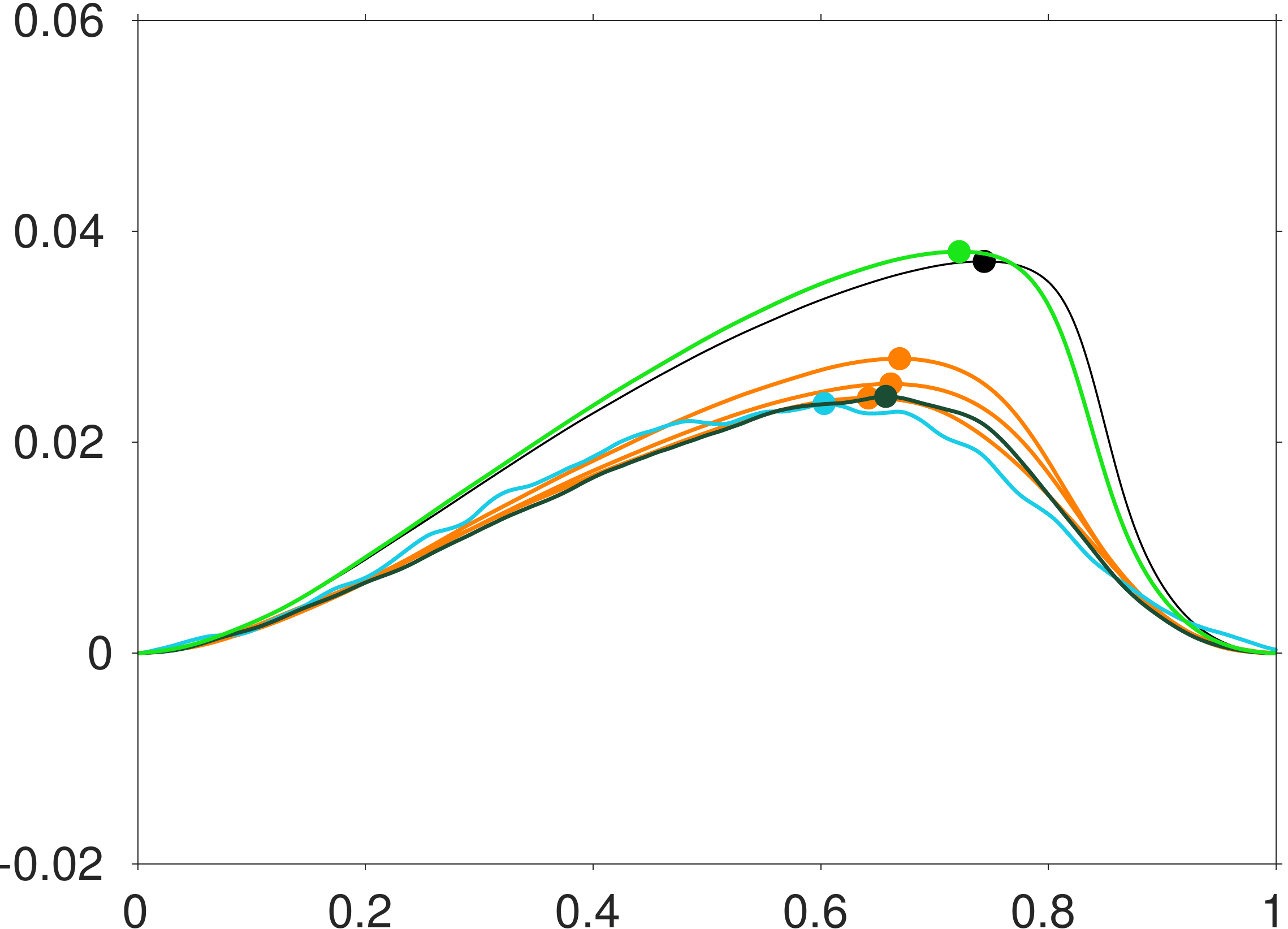}
           \centerline{\small $\tilde{x}/\lambda_h$}
         \end{minipage}
         \begin{minipage}{2ex}
           \rotatebox{90}
           {\small \hspace{4ex}$AR$}
         \end{minipage}
         \begin{minipage}{.45\linewidth}
           \raggedright{(\textit{b})}
           \includegraphics[width=\linewidth]
           {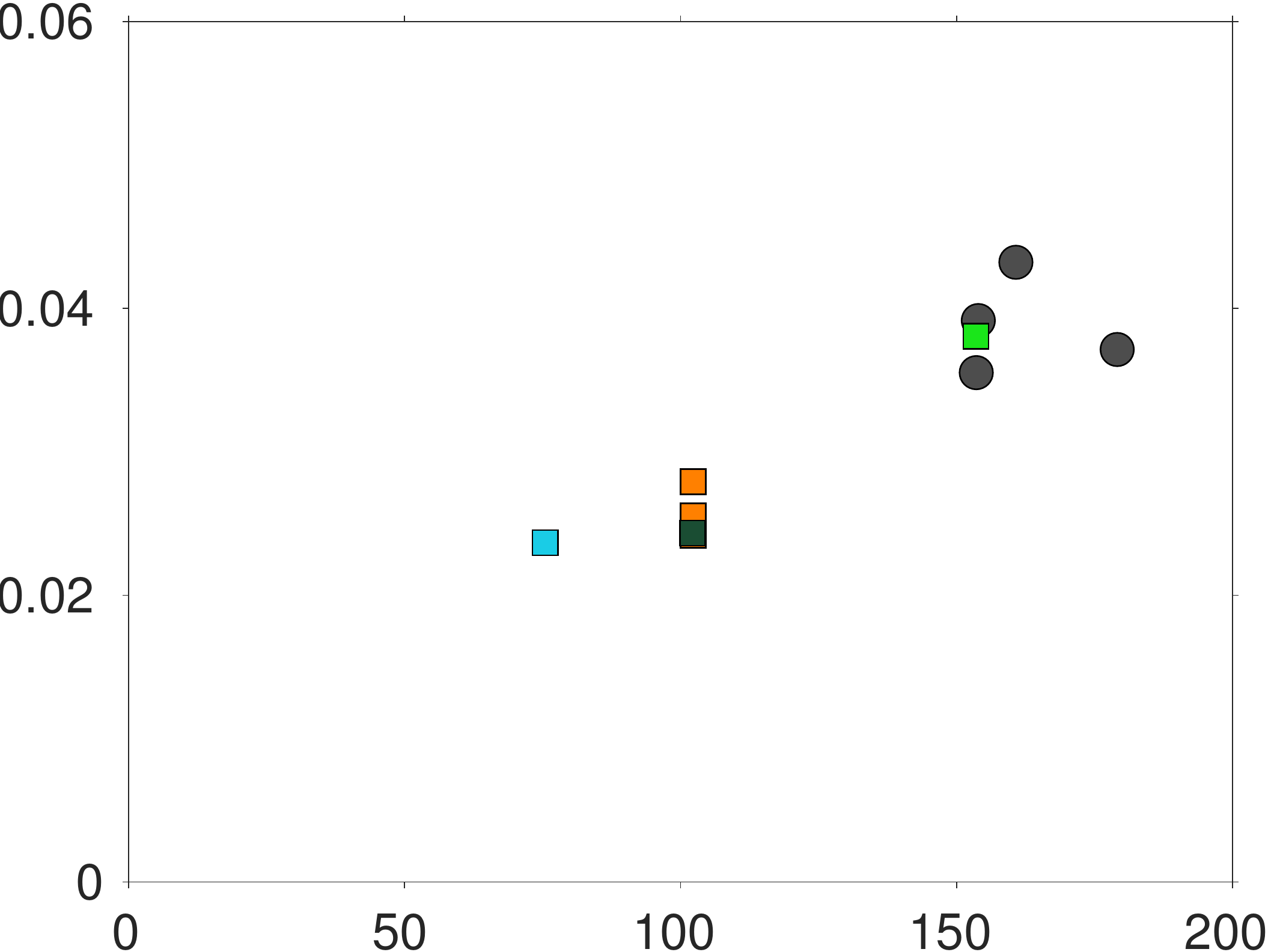}
           \centerline{\small $\lambda_h/D$}
         \end{minipage}
  	\caption{(\textit{a}) Mean two-dimensional profile of the sedimentary patterns, averaged over the time period $T^s_{obs}$ at the end of each simulation. The filled circles indicate the location, where each profile attains its maximal value. Note that for the sake of visualization, the vertical and the horizontal axes are shown in different scales, i.e. the aspect ratio is exaggerated. Cases
    \icaseC{} ({{\color{col_023} $\solidthick$}}),
    \icaseD{} ({{\color{col_120} $\solidthick$}}),
    \icaseE{} ({{\color{col_022} $\solidthick$}}),
    \icaseXabc{} ({\color{col_036} $\solidthick$}).
    The black line shows the same quantity for case $H7$ from \KU{}, with $L_x/D=179.2$ and $L_x/H_f=7.08$.
    (\textit{b}) Aspect ratio $AR$ of the mean two-dimensional profiles presented in (\textit{a}) as a function of the mean wavelength $\lambda_h$ (averaged over the final time interval $T^s_{obs}$). Round symbols represent case
    $H7$ as well as the three individual runs of case $H12$ from \KU{}.
    }
\label{fig:shape_mean_interf_steady}
\end{figure}

%
%
Figure \ref{fig:shape_mean_interf_steady} shows the mean pattern profile averaged over the final time interval of all simulations as well as the attained aspect ratio of the bedforms.
It can be seen, that the profiles of cases with $L_x/D \geq 153.6$ almost collapse upon a self-similar profile exhibiting a strong asymmetry ($LR \approx 0.28$) and an aspect ratio of $AR \approx 0.037$ (\KU{}).
Some experimental studies report higher values for the steady-state aspect ratio, i.e. $AR \approx 0.045$ \citep{Fourriere_al_2010}, $AR \approx 0.05$ \citep{Andreotti_Claudin_2013} and $AR \approx 0.067$ \citep{Charru_al_2013}. \KU{} explain their lower value with the fact, that in their cases, ``the `natural' steady-state regime is not yet reached'' and thus, a larger aspect ratio might be attained in a later phase, whereas the experimental data reflects the long-time state of the system which in those cases has evolved for at least one order of magnitude longer times.
On the other hand, the pattern profiles in cases \icaseD{} and \icaseXabc{} with $L_x = 102.4D$ are characterized by a smaller aspect ratio of $AR \approx 0.024 - 0.028$ and a more symmetric shape ($LR \approx 0.33 - 0.036$).
Important to note is, however, that all four profiles with $L_x = 102.4D$ are again nearly self-similar, despite the fact that $L_x/H_f$ varies by up to a factor of $1.5$. This is another indication that the pattern shape is mainly controlled by the particle diameter $D$, but almost unaffected by the mean fluid height $H_f$.
Further reducing the relative domain length to a value $L_x/D=76.8$ in case \icaseC{} leads to an even more symmetric profile ($LR \approx 0.41$), whereas the aspect ratio still attains a value $AR \approx 0.024$, comparable to the cases with $L_x = 102.4D$. An additional observation in case \icaseC{} is that, in contrast to the previous cases, the curve representing the pattern profile is not smooth, but exhibits wavy disturbances. This is a direct consequence of the lower amount of particles in case \icaseC{}, which leads to stronger disturbances of the spanwise-averaged pattern profile.

    \subsection{Variation of the mean fluid height}\label{sec:results_Hvar}
    In order to study the dependence of the minimal unstable wavelength $\lambda_{th}$ on the mean fluid height $H_f$, a second set of three simulations will be analysed in the following, including case \icaseXabc{} of \KU{} as well as two new simulations \icaseY{} and \icaseZ{}. All three cases have the same initial sediment bed configuration and the streamwise and spanwise box dimensions match. However, in case \icaseY{} and \icaseZ{}, $H_f$ has been increased by a factor of two compared to case \icaseXabc{}. Consequently, the bulk and friction Reynolds number are higher in the former two cases (cf. table~\ref{tab:param_phys}). In cases \icaseY{} and \icaseZ{}, two different values have been chosen for the Shields number, i.e. $\theta=0.18$ in the former and $\theta=0.13$ in the latter, by adjusting the Galileo number, while the remaining parameters are identical.
The dimensions of the channel configuration in the new simulations have been chosen in such a way, that their relative domain length ($L_x/H_f \approx 2$) lies below the lower bound for the most unstable wavelength as initially reported by \KU{}. As a consequence, none of the two cases should allow the sediment bed to become unstable and transverse patterns to evolve, if the minimal unstable wavelength scales with the mean fluid height. Note, however, that these authors did not vary $L_x/D$ and $L_x/H_f$ independently, and, therefore, could not distinguish between the two alternative scaling relations. Here, we are now in a position to do this.

\begin{figure}[t]
    \centering
         \begin{minipage}{2ex}
           \rotatebox{90}
           {\small \hspace{4ex}$ z/D$}
         \end{minipage}
         \begin{minipage}{.315\linewidth}
           \raggedright{(\textit{a})}
           \includegraphics[width=\linewidth]
           {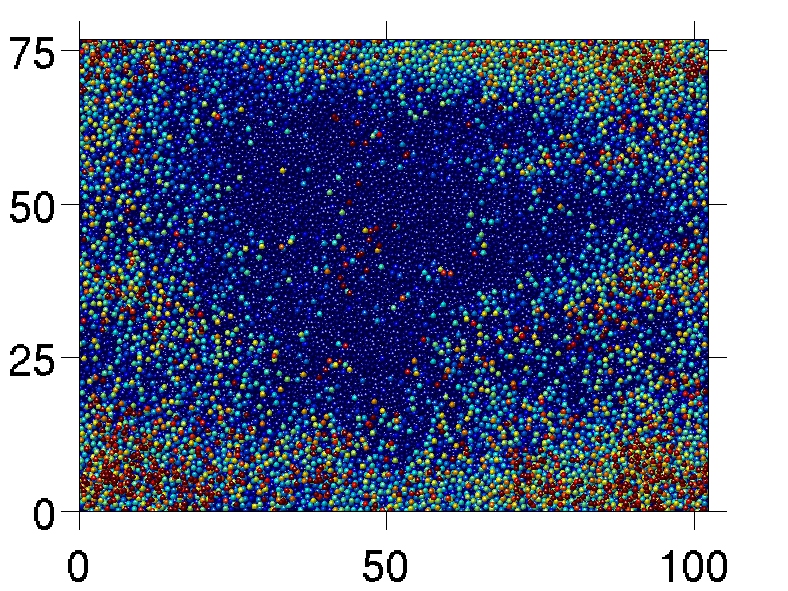}
           \centerline{\small $ x/D$}
         \end{minipage}
		 \begin{minipage}{.315\linewidth}
           \raggedright{(\textit{b})}
           \includegraphics[width=\linewidth]
           {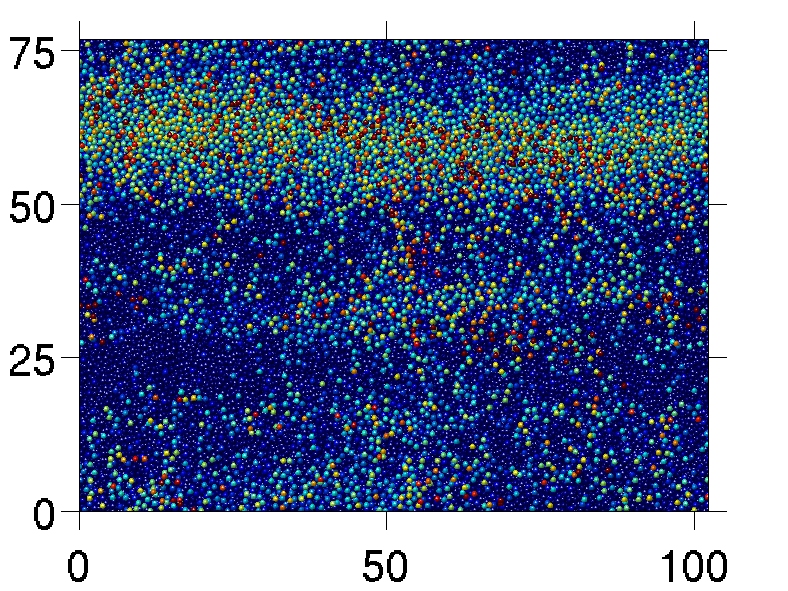}
           \centerline{\small $ x/D$}
         \end{minipage}
         \begin{minipage}{.315\linewidth}
           \raggedright{(\textit{c})}
           \includegraphics[width=\linewidth]
           {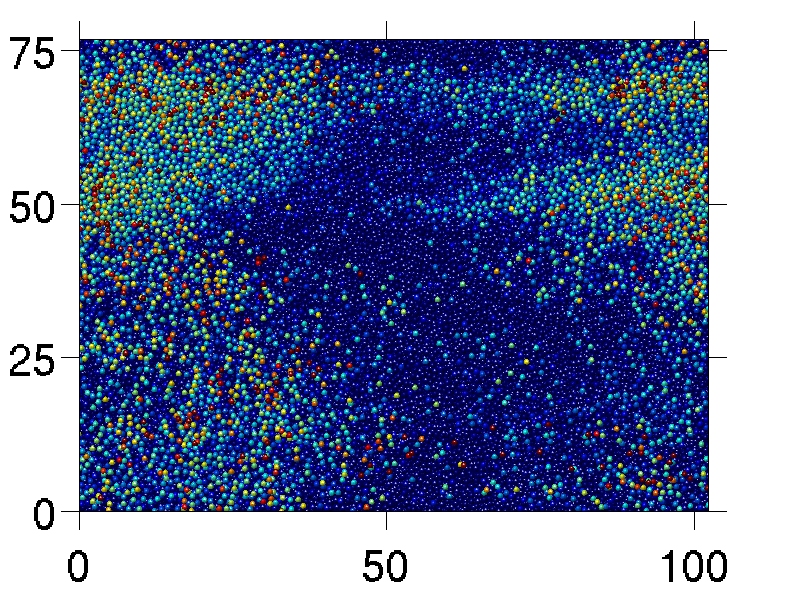}
           \centerline{\small $ x/D$}
         \end{minipage}
\caption{Instantaneous snapshots of the sediment bed of case \icaseY{}
        at (\textit{a})~{$t \approx 471 T_b$},
           (\textit{b})~{$t \approx 619 T_b$} and
           (\textit{c})~{$t \approx 687 T_b$},
           seen from the top of the channel. Colouring is the same as in figure \ref{fig:part_snapshots_D}.
}
\label{fig:part_snapshots_case100}
\centerline
    \centering
         \begin{minipage}{2ex}
           \rotatebox{90}
           {\small \hspace{4ex}$ z/D$}
         \end{minipage}
         \begin{minipage}{.315\linewidth}
           \raggedright{(\textit{a})}
           \includegraphics[width=\linewidth]
           {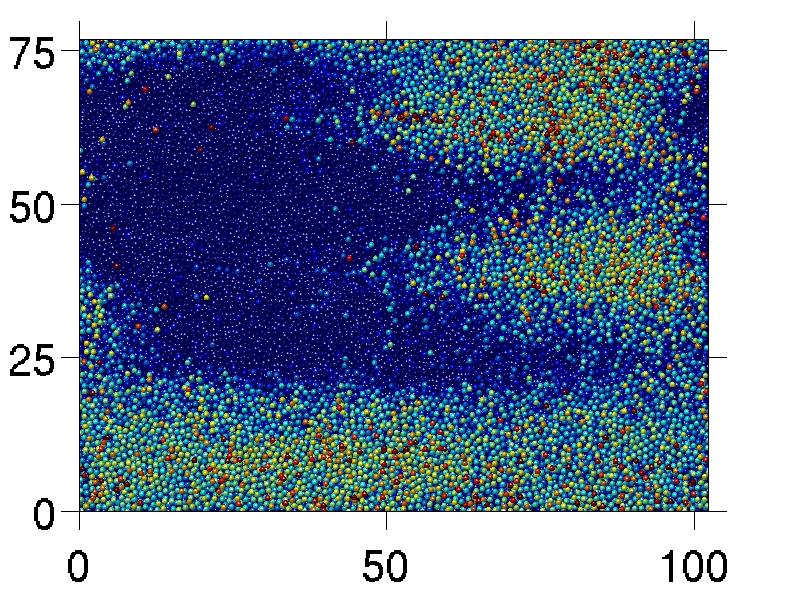}
           \centerline{\small $ x/D$}
         \end{minipage}
		 \begin{minipage}{.315\linewidth}
           \raggedright{(\textit{b})}
           \includegraphics[width=\linewidth]
           {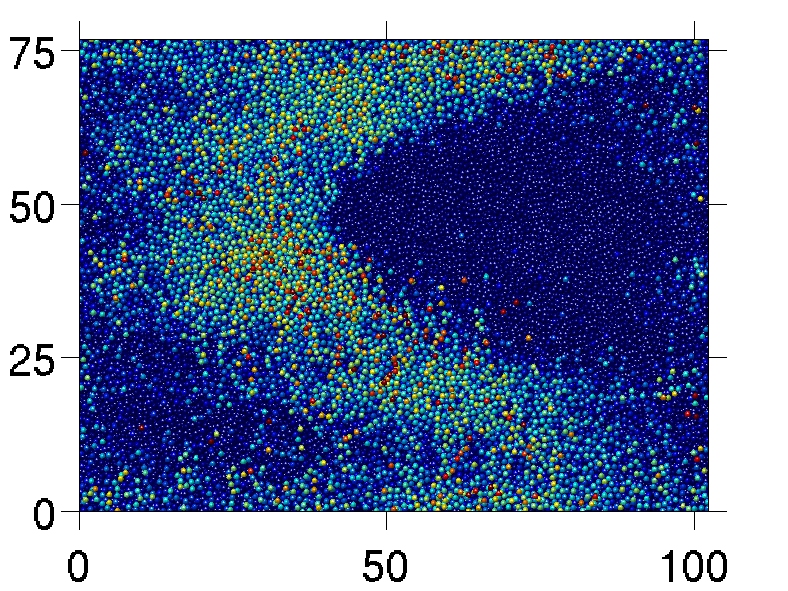}
           \centerline{\small $ x/D$}
         \end{minipage}
         \begin{minipage}{.315\linewidth}
           \raggedright{(\textit{c})}
           \includegraphics[width=\linewidth]
           {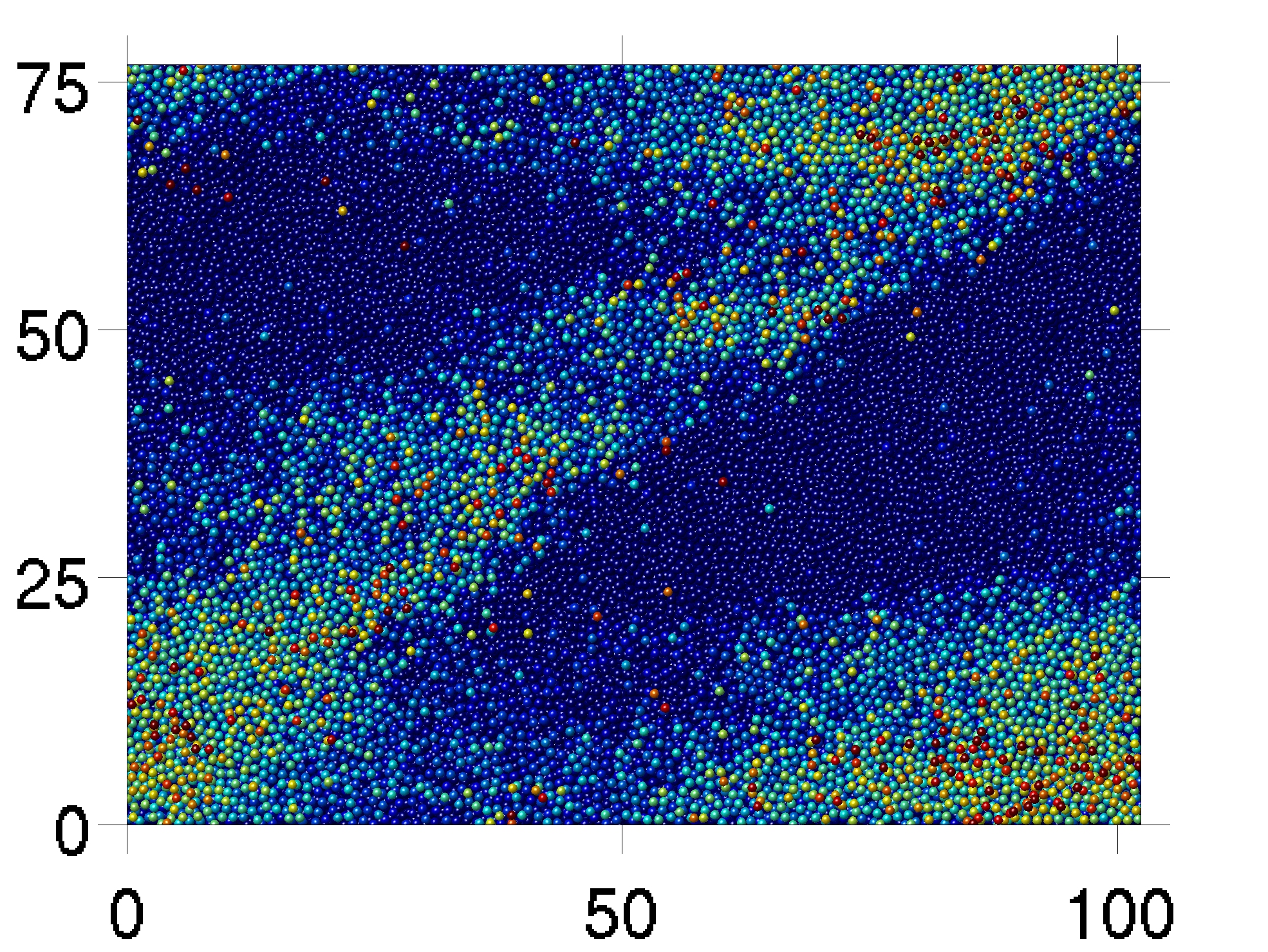}
           \centerline{\small $ x/D$}
         \end{minipage}
\caption{Instantaneous snapshots of the sediment bed of case \icaseZ{} 
        at
        (\textit{a})~{$t \approx 300 T_b$},
        (\textit{b})~{$t \approx 352 T_b$} and
        (\textit{c})~{$t \approx 490 T_b$},
        seen from the top of the channel. Colouring is the same as in figure \ref{fig:part_snapshots_D}.
}
\label{fig:part_snapshots_case101}
\end{figure}

%
%
%
Figure \ref{fig:part_snapshots_case100} and figure \ref{fig:part_snapshots_case101} show selected instantaneous snapshots of the particle bed evolution for the new cases \icaseY{} and \icaseZ{}, respectively.
The sediment bed surface of case \icaseY{} is deformed showing streamwise and spanwise elongated crestlines. In some phases, the system alters between patterns at either of the two orientations (cf. figure \ref{fig:part_snapshots_case100}(\textit{b}) and \ref{fig:part_snapshots_case100}(\textit{c})),
whereas in figure \ref{fig:part_snapshots_case100}(\textit{a}), for instance, streamwise and spanwise oriented crestlines appear at the same time.
Similarly, the sediment bed of case \icaseZ{} exhibits one crest line parallel and one perpendicular to the mean flow direction in figure \ref{fig:part_snapshots_case101}(\textit{a}). In contrast, a later snapshot in figure   \ref{fig:part_snapshots_case101}(\textit{b}) shows a three-dimensional bedform with a horse-shoe like shape with horns on both sides pointing downstream. This pattern resembles in its geometry a barchan dune \citep{Franklin_Charru_2011}. In the last snapshot provided in figure \ref{fig:part_snapshots_case101}(\textit{c}) a diagonal crest line, crossing the whole extent of the $x$-$z$-plane from the lower left to the upper right corner, has evolved.
These observations represent a marked difference between the systems with larger clear fluid height ($H_f/D \approx 50$) and those with smaller relative submergences ($H_f/D \approx 25$). In the latter case (simulations \icaseXabc{} of \KU{} - images not shown), regular streamwise-aligned ridges are either displaced by the higher, dominating transverse bedforms or do exist superimposed to the former ones.

An important consequence of the observed behaviour in cases \icaseY{} and \icaseZ{} is that the bed height is clearly two-dimensional.
This means that the spanwise-averaged sediment bed and flow field will not correctly describe the physical processes that lead to the formation of these type of bedforms. In the remainder of the current work, we will therefore analyse the sediment bed in its entire streamwise and spanwise dimension and omit the spanwise averaging (cf. the definition in section~\ref{subsec:2D_interface}).

\begin{figure}[t]
    \centering
         \begin{minipage}{3ex}
           \rotatebox{90}
           {\small \hspace{2ex} $\sigma_h^{2D} /D$}
         \end{minipage}
         \begin{minipage}{.475\linewidth}
           \raggedright{(\textit{a})}
           \includegraphics[width=\linewidth]
           {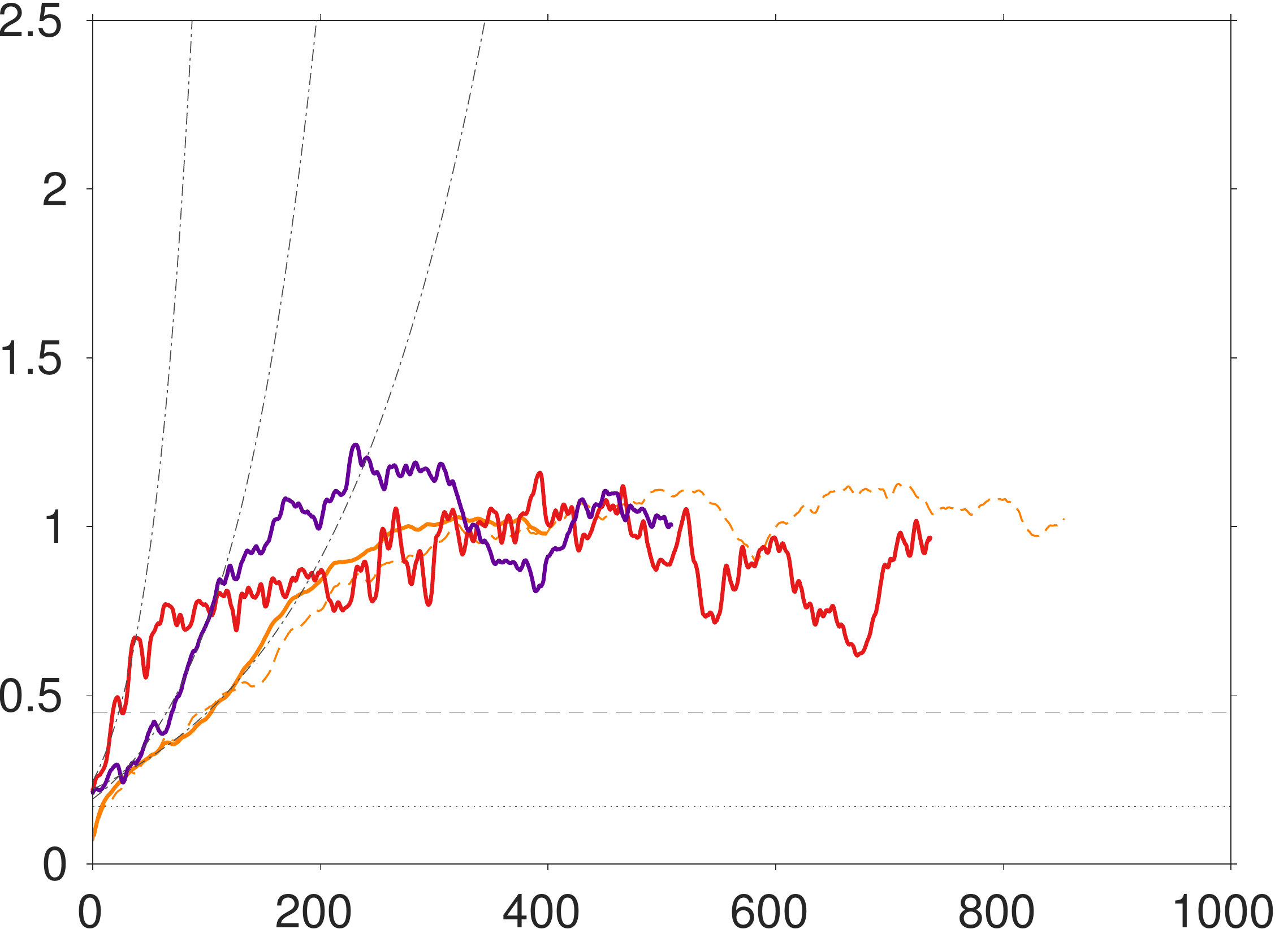}
           \centerline{\small $ t/(H_f/u_b)$}
         \end{minipage}
         \begin{minipage}{.475\linewidth}
           \raggedright{(\textit{b})}
           \includegraphics[width=\linewidth]
           {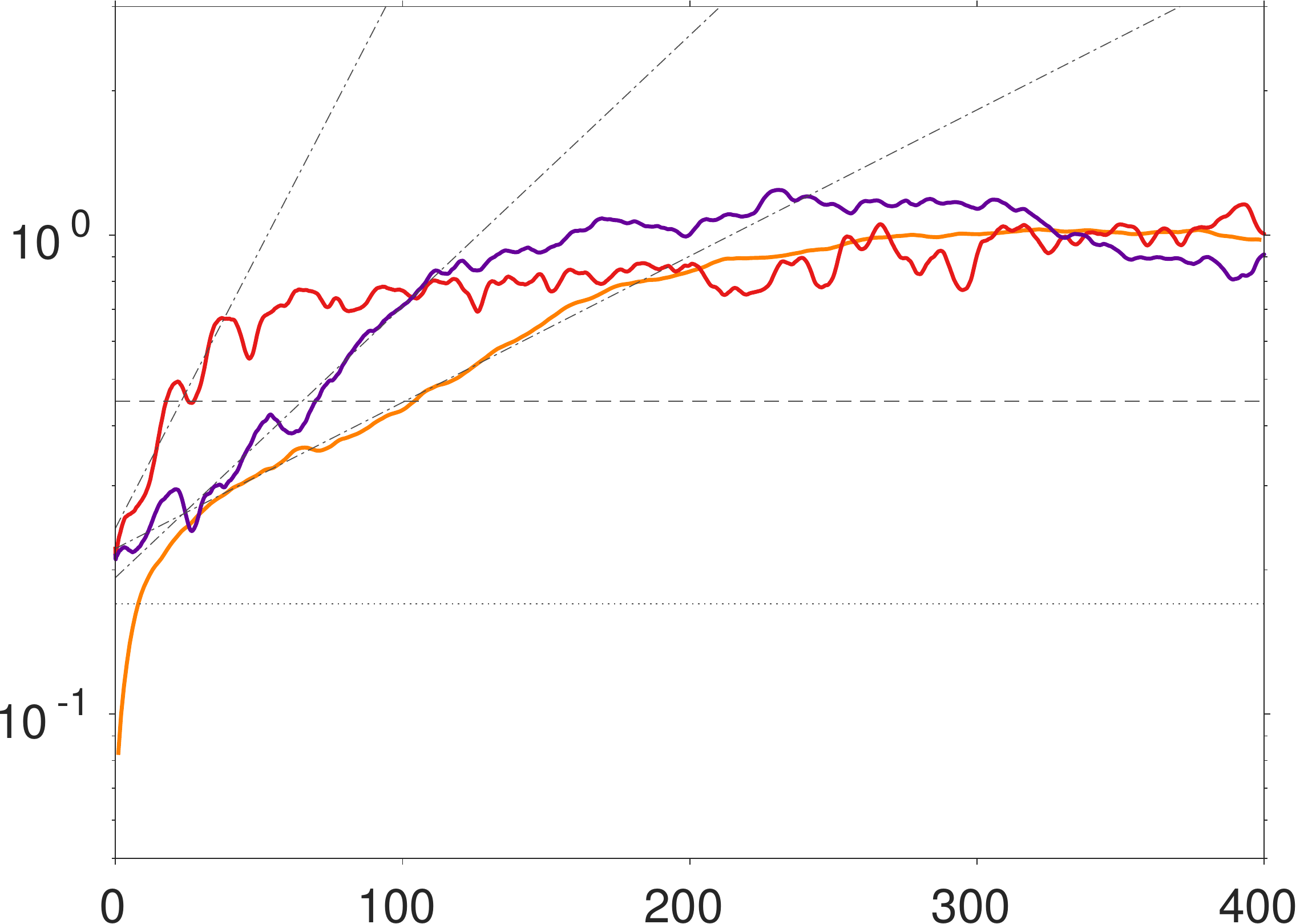}
           \centerline{\small $ t/(H_f/u_b)$}
         \end{minipage}
  	\caption{
    (\textit{a}) Time evolution of the two dimensional root mean square of the bedform amplitude normalized by the particle diameter $\sigma_h^{2D}/D$. Time is scaled in bulk time units $T_b$.
    Cases
    \icaseXabc{} ({{\color{col_034} $\solidthick$}}),
    \icaseY{} ({{\color{col_100} $\solidthick$}}),
    \icaseZ{} ({{\color{col_101} $\solidthick$}}).
    The data of case \icaseXabc{} is presented as ensemble average over the three simulations ({{\color{col_034} $\solidthick$}}, thick line). Additionaly, the individual evolution of run \icaseXc{} is presented ({{\color{col_034} $\dashed$}}, thin line).
    The dashed-dotted lines represent exponential curves of the form $\sigma_h/D = A\exp(Bt/T_b)$ which have been found to best fit the evolution of $\sigma_h^{2D}$ during the initial growth period of the respective case. The coefficients are as follows:
    \icaseXabc{}: $A=0.2215$, $B=0.0070$;
    \icaseY{}: $A=0.2441$, $B=0.0266$;
    \icaseZ{}: $A=0.1926$, $B=0.0130$.
    Note that in case \icaseXabc{}, the exponential curve best fits the ensemble average over the three runs.
    (\textit{b}) Same data as (\textit{a}), but represented in semi-logarithmical scale.
    }
\label{fig:overview_rms2D_bedheight_evolution}
\end{figure}

%
%
Figure~\ref{fig:overview_rms2D_bedheight_evolution} shows the time evolution of the root mean square of the two-dimensional sediment bed height fluctuation $\sigma_h^{2D}$.
All cases exhibit an exponential growth interval, but with clearly different growth rates, as indicated by the fitted exponential curves presented in figure~\ref{fig:overview_rms2D_bedheight_evolution}.
It is important to keep in mind that in contrast to the spanwise-averaged case, the two-dimensional measure $\sigma_h^{2D}$ takes into account streamwise and spanwise variations of the fluid-bed interface, which means that a change in $\sigma_h^{2D}$ with time can be a sign of evolving ridges or transverse patterns likewise.
Indeed, the detailed analysis of a two-dimensional Fourier decomposition of the bed height perturbation which is presented below shows that the amplitude of variations in both directions is of similar order.
In order to determine the scaling of the initial growth rate of the sediment bed height fluctuation, however, one would require a larger database and a larger number of individual runs at each parameter point to allow for ensemble averaging similar to case \icaseXabc{}. Nevertheless, it would be of high interest to elucidate the influence of the relevant parameters on the initial growth rate of $\sigma_h^{2D}$ such as the Reynolds number $Re_b$ ($Re_{\tau}$, respectively) and the relative submergence $H_f/D$ in a future study.
After roughly $300$ bulk time units, all cases eventually reach what could be called an asymptotic state. We observe that the pattern amplitude attains fully-developed averaged values of similar order comparable to one particle diameter for all cases, which suggests that at the given parameter point shown in figure \ref{fig:overview_rms2D_bedheight_evolution}, the averaged value of $\sigma_h^{2D}$ in the final interval does not strongly depend on the value of the mean fluid height $H_f$.

%
%
The analysis of $\sigma_h^{2D}$ has shown that the sediment bed becomes unstable in all three cases. However, as $\sigma_h^{2D}$ is an integral measure for the bed evolution, it cannot provide further information about the evolution and interaction of single unstable modes. In particular, it does not allow to determine the role of streamwise and transverse modes separately. For this reason, the evolution of the bed height perturbation will be further analysed in Fourier space for the different streamwise and spanwise wavenumbers. To this end, we compute the single-sided amplitude spectra $\hat{A}_{(k,l)}$ (for the physically relevant non-negative wavenumbers $k,l \geq 0$) as twice the absolute value of the coefficient $\hat{h}_{b(k,l)} (\kappa^1_k,\kappa^3_l,t)$, which is the Discrete Fourier Transform (DFT) of the sediment bed height perturbation in the physical space, $h_b' (x,z,t)$. In the above expression, $\kappa^d_k$ is the wavenumber of the $k$-th mode in spatial direction $d$ (where $d=1,3$ corresponds to the $x$- and $z$-direction, respectively).

\begin{figure}[t!]
    \centering
         \begin{minipage}{2ex}
           \rotatebox{90}
           {\small \hspace{5ex}$\hat{A}_{(k,l)}/D$}
         \end{minipage}
         \begin{minipage}{.46\linewidth}
		       \raggedright{(\textit{a})}
           \includegraphics[width=\linewidth]
           {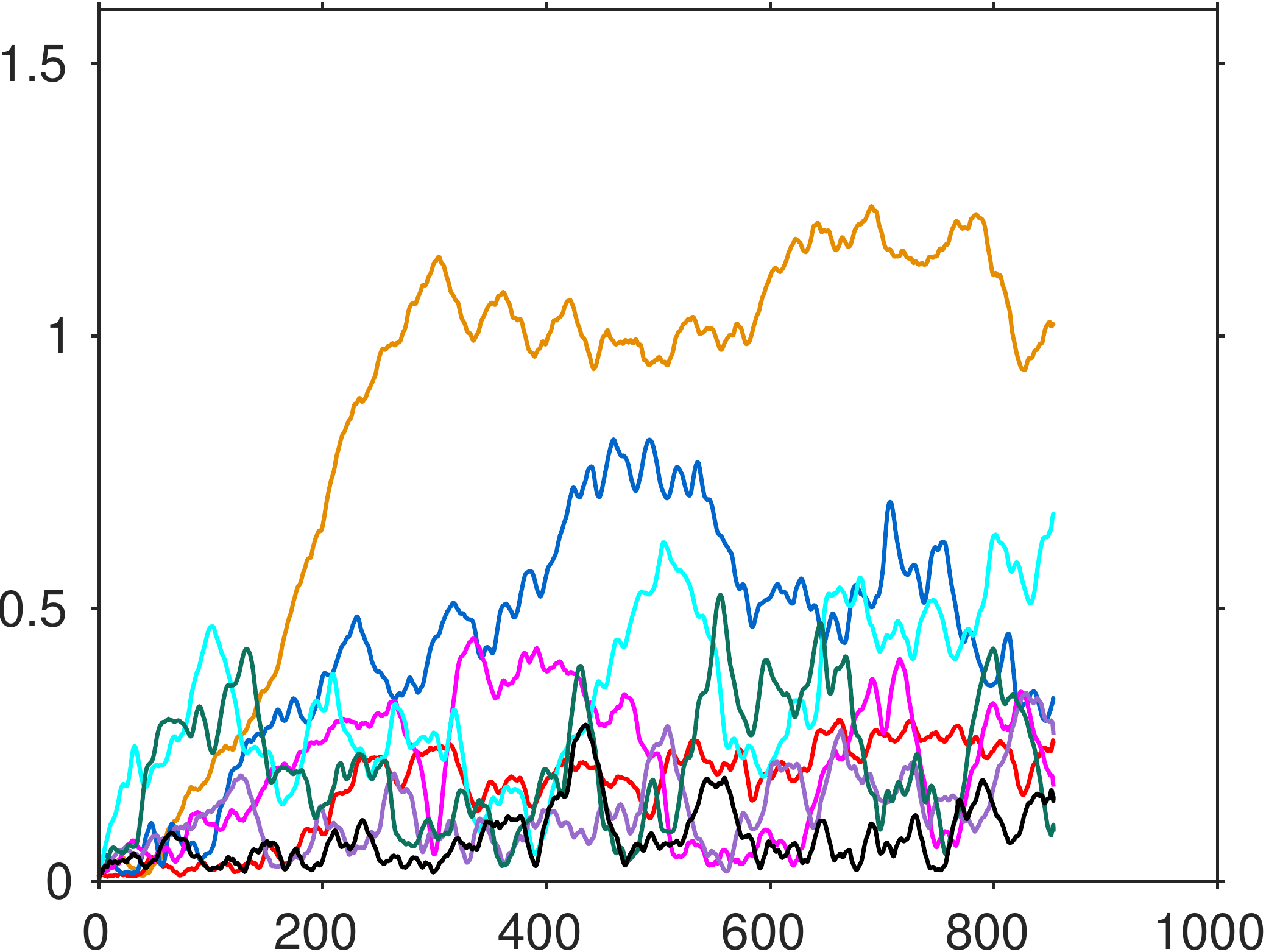}
           \centerline{\small $ t/(H_f/u_b)$}
         \end{minipage}
         \hfill
         { }\\[3ex]
         \begin{minipage}{2ex}
           \rotatebox{90}
           {\small \hspace{5ex}$\hat{A}_{(k,l)}/D$}
         \end{minipage}
         \begin{minipage}{.46\linewidth}
           \raggedright{(\textit{b})}
		       \includegraphics[width=\linewidth]
           {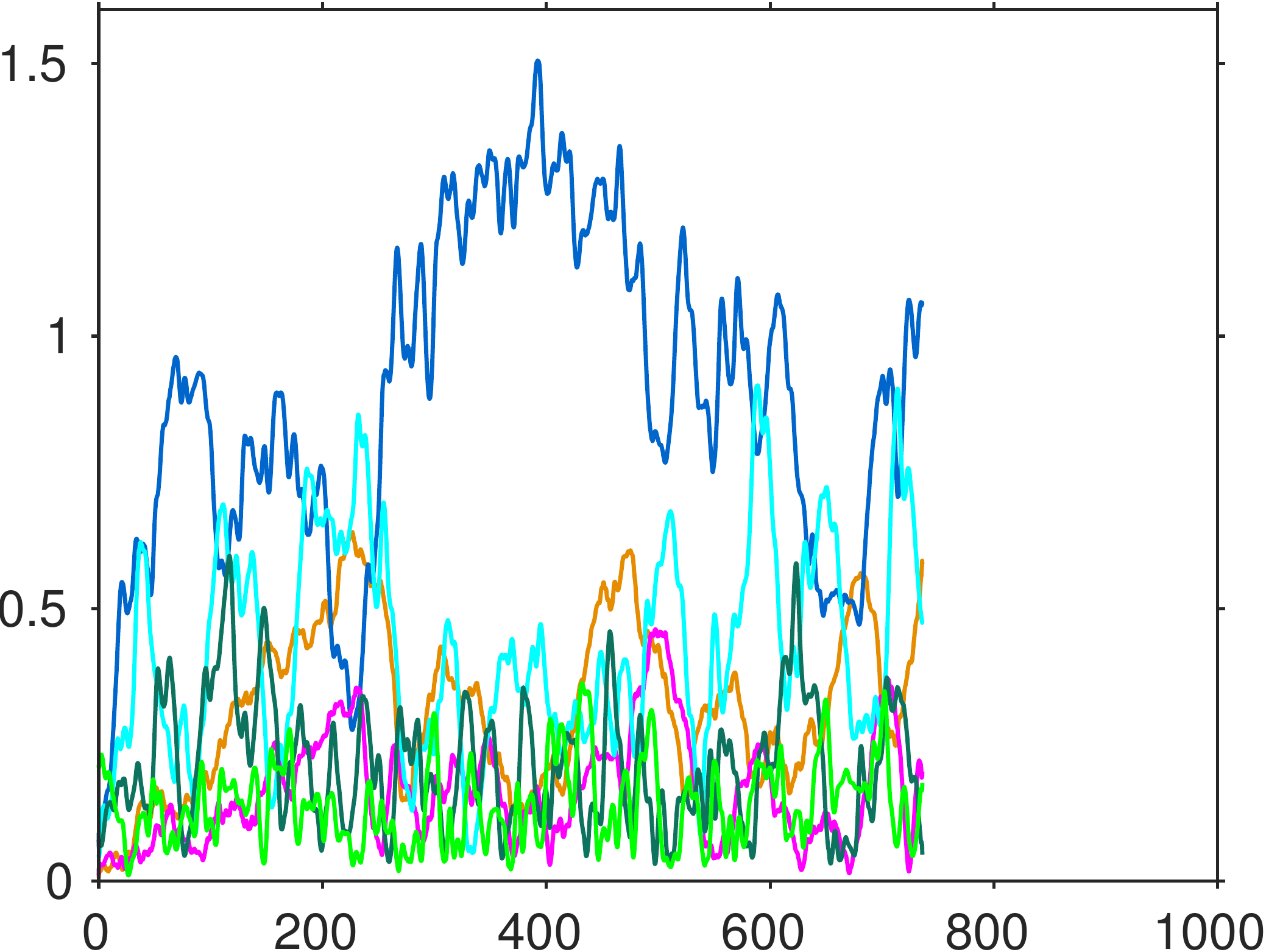}
           \centerline{\small $ t/(H_f/u_b)$}
         \end{minipage}
         \begin{minipage}{.46\linewidth}
		       \raggedright{(\textit{c})}
           \includegraphics[width=\linewidth]
           {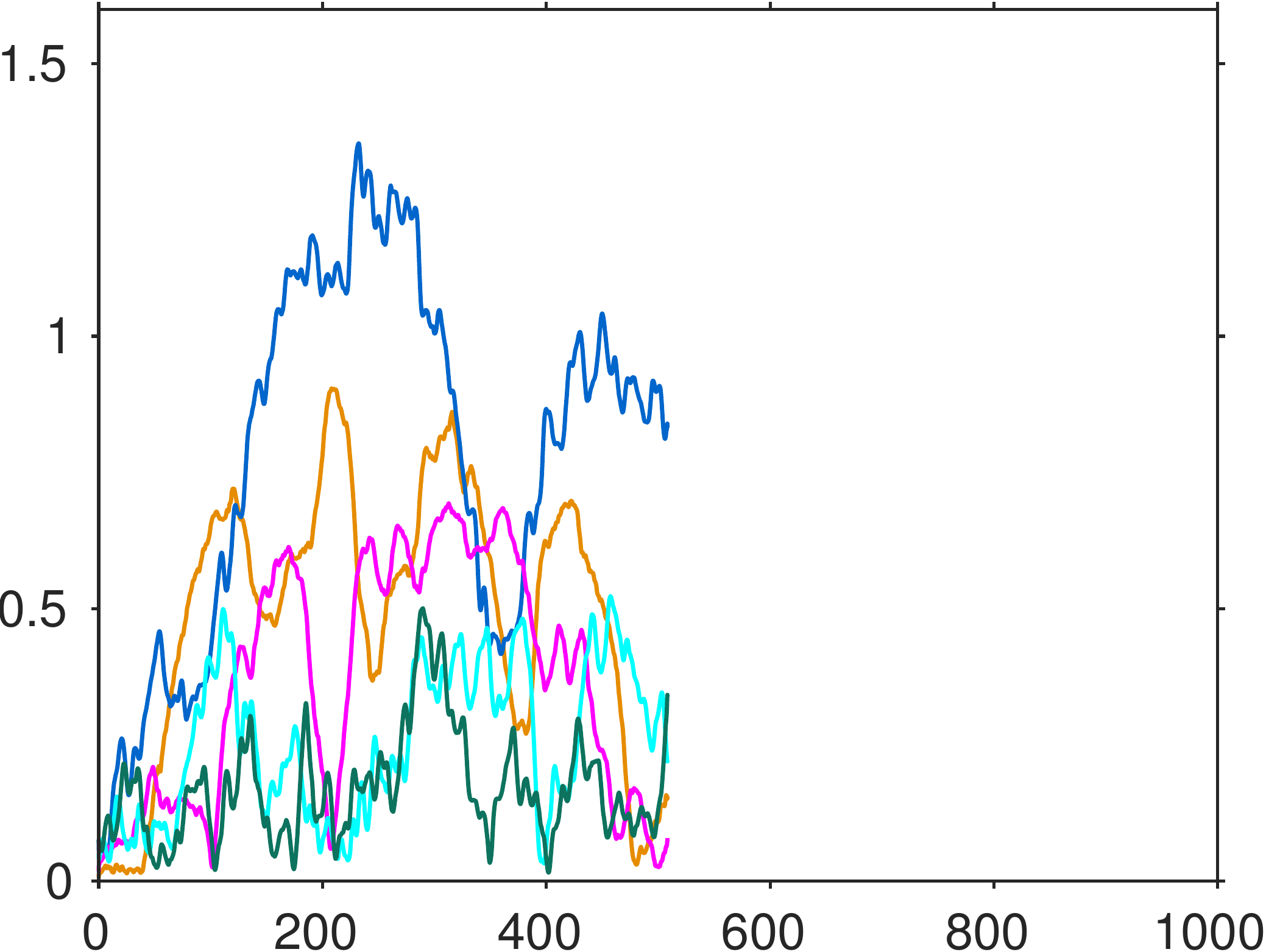}
           \centerline{\small $ t/(H_f/u_b)$}
         \end{minipage}

  	\caption{Time evolution of the single-sided amplitude spectra for the most dominant modes
	  normalized with the particle diameter $\hat{A}_{(k,l)}/D$ for cases
    (\textit{a}) \icaseXc{},
  	(\textit{b}) \icaseY{} and
  	(\textit{c}) \icaseZ{}.
	  Note that only those modes are defined as dominant, that exceed a value of
    $\hat{A}_{(k,l)} = 0.30D$ during the simulation. The notation $\hat{A}_{(k,l)}$
    indicates that the amplitude corresponds to the mode with streamwise and spanwise
    wavenumber $\kappa^1_k$ and $\kappa^3_l$, respectively.
  	Colouring of the individual dominant modes is as follows:
  	{{\color{colmode_1_0} $\solidthick$}}, $\hat{A}_{(1,0)}$;
  	{{\color{colmode_2_0} $\solidthick$}}, $\hat{A}_{(2,0)}$;
  	{{\color{colmode_0_1} $\solidthick$}}, $\hat{A}_{(0,1)}$;
  	{{\color{colmode_0_2} $\solidthick$}}, $\hat{A}_{(0,2)}$;
  	{{\color{colmode_0_3} $\solidthick$}}, $\hat{A}_{(0,3)}$;
  	{{\color{colmode_0_4} $\solidthick$}}, $\hat{A}_{(0,4)}$;
  	{{\color{colmode_1_1} $\solidthick$}}, $\hat{A}_{(1,1)}$;
  	{{\color{colmode_1_2} $\solidthick$}}, $\hat{A}_{(1,2)}$;
    {{\color{colmode_1_3} $\solidthick$}}, $\hat{A}_{(1,3)}$.
}
\label{fig:2D_amplitude_spectra_evo}
\end{figure}

%
%
Figure \ref{fig:2D_amplitude_spectra_evo} shows the time evolution of the single-sided amplitude spectra for the most dominant modes in case  \icaseXc{}, \icaseY{} and \icaseZ{}, respectively.
In case \icaseXc{}, the wave with modes $(1,0)$ (first harmonic in the streamwise, constant in the spanwise direction) attains the highest amplitudes and thus clearly dominates the spectra. In good agreement with the evolution of $\sigma_h^{2D}$, this mode first increases exponentially during the initial $200-250$ bulk time units, before it settles at values which fluctuate somewhat around an asymptotic mean.
Note that only the first two harmonics of the pure streamwise waves $\lambda_{(1,0)}$ and $\lambda_{(2,0)}$ exceed a value of $\hat{A}_{(k,l)} = 0.30D$ during the observation interval, highlighting that in these cases with a comparably short domain, a very sparse range of wavenumbers is amplified, and is then available to form the sediment pattern.
On the other hand, waves of the first three pure spanwise harmonics $\lambda_{(0,1 \ldots 3)}$ are found to be significant.
In the first approximately $100-150$ bulk time units, these pure spanwise modes dominate the amplitude spectra, which reflects the formation of initial streamwise aligned ridges that has been observed by \KU{}. In the subsequent quasi-steady phase of the bedform, pure spanwise modes are still present with finite but smaller amplitudes compared to $\hat{A}_{(1,0)}$, reaching maximum values $\hat{A}_{(0,1)} = 0.80D$. This is in good agreement with the observation of the aforementioned authors, that streamwise elongated patterns are still visible on the upstream face of the transverse bedforms, once these latter have reached a quasi-steady state. In addition, several diagonal waves including modes larger than zero in both directions evolve, but they do not reach amplitudes higher than  $\hat{A} = 0.5D$.
The time evolution of the amplitude spectra in the new simulations \icaseY{} and \icaseZ{} (cf. figures \ref{fig:2D_amplitude_spectra_evo}(\textit{b}),(\textit{c}), respectively), on the other hand, is dominated by the first pure spanwise oscillating wave mode $(0,1)$ during almost the whole observation interval, attaining maximum values of $\hat{A}_{(0,1)}/D$ above unity.
It is worth to note that this latter is the only pure streamwise wave which is significant.
In general, the amplitude of the single modes exhibit higher fluctuations than those in case \icaseXc{}, leading to differences of approximately $1D$ between the highest and the lowest attained values of a single mode. During the entire simulation, none of the modes reach a plateau regime which would indicate a quasi-steady state of the bed. These findings match our observations, that both systems \icaseY{} and \icaseZ{} show a sequence of different alternating bedform configurations composed of a number of streamwise and spanwise unstable modes without eventually reaching a quasi-steady state.
In particular, transverse sediment bed waves evolve and, even though they are observed to be of smaller amplitude than their spanwise directed counterparts, they are involved in the formation of three dimensional patterns. Eventually, this reveals that, indeed, a sediment bed can become unstable for a domain length $L_x/H_f < 3$, indicating that the lower bound for the most-unstable wavelength as reported by \KU{} does not scale with the mean fluid height $H_f$. The three-dimensional pattern evolution, however, requires further investigation.

  \section{Discussion}\label{sec:discussion}
\begin{figure}[t!]
    \centering
         \begin{minipage}{2ex}
           \rotatebox{90}
           {\small \hspace{6ex}$L_x/H_f, \lambda/H_f$}
         \end{minipage}
         \begin{minipage}{.70\linewidth}
           \centerline{}
           \includegraphics[width=\linewidth]
           {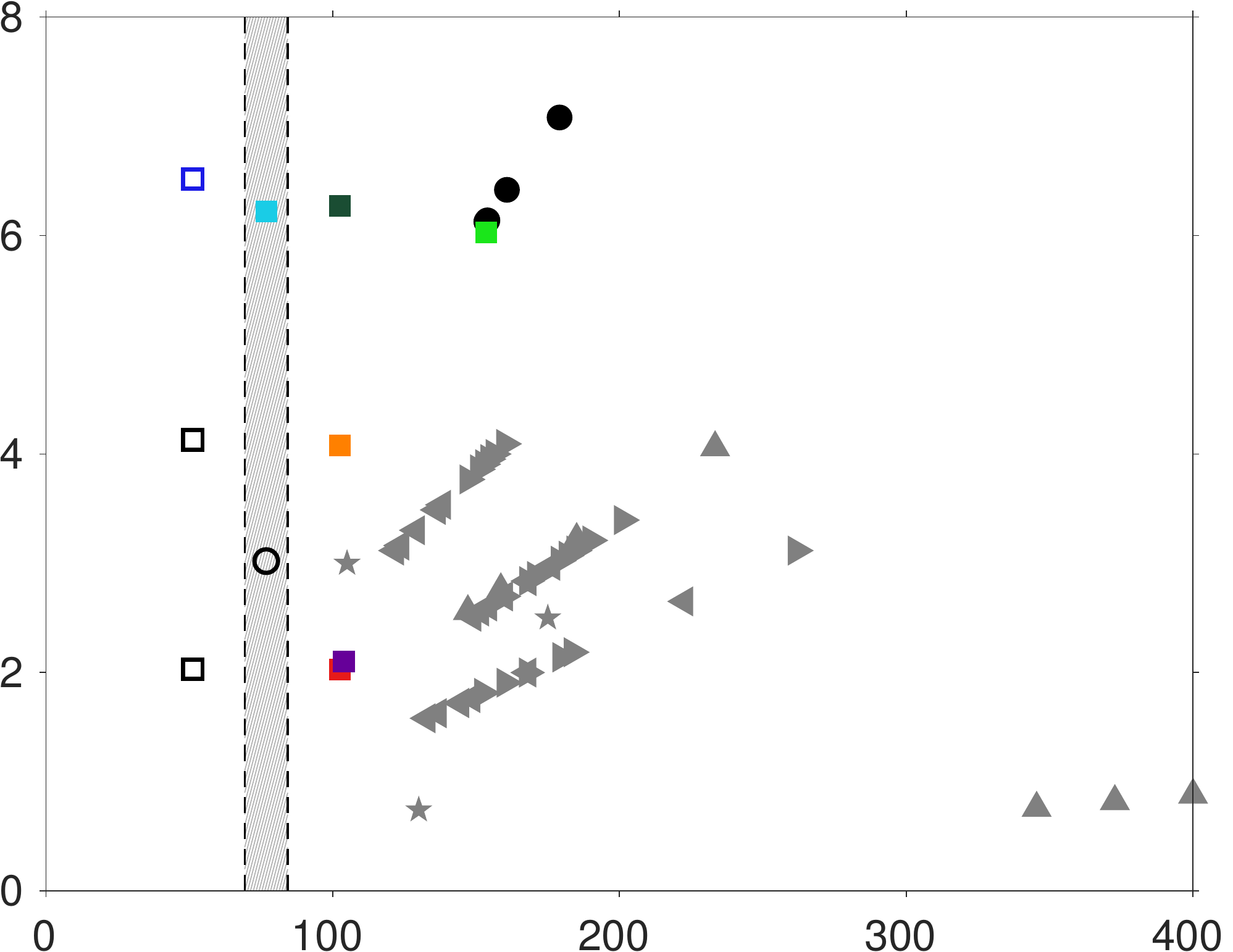}
           \centerline{\small $L_x/D, \lambda/D$}
         \end{minipage}
    \caption{Minimal unstable wavelengths in numerical simuations and experiments as functions
    of the particle diameter $D$ and the mean fluid height $H_f$.
    Filled symbols indicate unstable sediment beds, whereas open symbols are used for stable systems.
    Round black symbols are data points from \KU{} {({\color{black} \solidcircle}, {\color{black} \opencircle})},
    while the simulations analysed in the current study are indicated as:
    \icaseAA{} ({\color{black} \opensquare}),
    \icaseA{} ({\color{black} \opensquare}),
    \icaseB{} ({\color{col_110} \opensquare}),
    \icaseC{} ({\color{col_023} \solidsquare}),
    \icaseD{} ({\color{col_120} \solidsquare}),
    \icaseE{} ({\color{col_022} \solidsquare}),
    \icaseXabc{} ({\color{col_036} \solidsquare}),
    \icaseY{} ({\color{col_100} \solidsquare}),
    \icaseZ{} ({\color{col_101} \solidsquare}).
    The presented wavelengths are determined as the time average of the mean wavelength
    $\lambda_h$ over the final time period $T^s_{obs}$.
    It should be noted that for stable systems as well as for cases \icaseY{} and \icaseZ{},
    in which three-dimensional patterns evolve, the streamwise boxlength is given
    instead of the spanwise-averaged mean wavelength (cf. the detailed discussion in the text).
    Wavelengths measured in experiments are presented with the following symbols:
    \cite{Coleman_al_2003} (closed-conduit, {\color{col_gray} \solidtriup}),
    \citet{Langlois_Valance_2007} (channel, {\color{col_gray} \solidstar}),
    \cite{CardonaFlorez_Franklin_2016} (closed-conduit, {\color{col_gray} \solidtrileft} for the first and {\color{col_gray} \solidtriright} for the last measured wavelength).
    Note that in the experimental studies, no free surface is present.
    Thus, the experimentally determined wavelengths
    are normalized with the half mean fluid height.
    Based on the results of our analysis, we expect the minimal unstable wavelength
    in the vertical grey region around ${\lambda/D = 80}$.
    }
  	\label{fig:overview_LxH_ov_LxD}
\end{figure}

%
%
Our observations in the previous section suggest that at the investigated parameter point, the minimal unstable wavelength depends on the particle diameter, whereas it seems to be rather unaffected by variations of the mean fluid height. In the following, we will compare our results with measurements as well as proposed scaling relations from experimental and theoretical studies.

In figure~\ref{fig:overview_LxH_ov_LxD}, we present an overview of initial mean pattern wavelengths from our current and previous direct numerical simulation studies together with values determined in laboratory experiments. The observed wavelengths are shown as functions of the particle diameter $D$ and of the mean fluid height $H_f$, respectively.
Note that for stable sediment beds (indicated by open symbols in figure~\ref{fig:overview_LxH_ov_LxD}), the definition of an initial pattern wavelength is meaningless. Here, we give instead the streamwise domain length $L_x$, which indicates the maximum possible wavelength that did not evolve during the simulation.
In cases \icaseY{} and \icaseZ{}, on the other hand, we have observed the sediment bed to evolve both in the streamwise and spanwise direction.
In these two cases we equally associate the value of the domain length $L_x$ with the most unstable wavelength, since the study of the single-sided amplitude spectra of the sediment bed height fluctuation shows that the most dominant streamwise mode is the one with indices $(1,0)$, i.e. the first streamwise harmonic which is constant in the spanwise direction.‚
It should be further mentioned that with cases \icaseAA{} and \icaseA{}, two additional new cases have been added to the parameter plane, which were not part of the analysis in the previous section (cf. table~\ref{tab:param_phys} and table~\ref{tab:param_numer} for the physical and numerical parameters). In both cases, however, the bed is observed to remain stable, i.e. no transverse patterns form during the simulation.
The experimental data shown in figure~\ref{fig:overview_LxH_ov_LxD} comes from measurements in channel flow by \citet{Langlois_Valance_2007} as well as in closed-conduit flows by \citet{Coleman_al_2003} and
\citet{CardonaFlorez_Franklin_2016}.

In our direct numerical simulations, we have observed transverse pattern formation only in cases with a relative streamwise domain length $L_x/D \geq 76.8$, while the sediment bed remained stable for all simulations below this limit, irrespective of the fluid height. At a domain length $L_x/D = 76.8$, we have observed case \icaseC{} to become unstable, but the shape and asymmetry of the evolving pattern is seen to substantially deviate from the observed values in sufficiently long domains such as the cases of \KU{} with a box length of $L_x/D = \mathcal{O}(10^3)$.
A possible reason for this observation is that in \icaseC{}, the system can only choose between 8 harmonics with a wavelength higher than $\mathcal{O}(D)$. The interaction of this small number of discrete modes may not lead to the same interface that would form in a sufficiently long domain.
Interestingly, \KU{} report a case with the same relative domain length $L_x/D = 76.8$ in which the sediment bed remained flat. We therefore expect that a box length $L_x/D = 76.8$ is in the vicinity of the sought threshold wavelength, such that at this parameter point, already small differences in the configuration might tip the system in either way. If the domain length is even further decreased (as in cases \icaseAA{}, \icaseA{} and \icaseB{}) the bed remains flat.

The existence of such a lower threshold for the most amplified wavelength originates in the interaction of several stabilizing and destabilizing effects in the flow system, which allow only a certain range of wavelengths to evolve. In order to show that this is indeed the case and that the absence of patterns in cases \icaseAA{}, \icaseA{} and \icaseB{} is not due to purely geometrical constraints, we have estimated the height of a pattern with wavelength $\lambda/D = L_x/D=51.2$ and a downstream face inclined at the angle of repose, which represents the highest possible pattern for a given domain length. On the other hand, a pattern at the same wavelength in the asymptotic state (i.e. with similar aspect ratio and degree of asymmetry as observed in the cases with $L_x/D \geq 150D$) would attain a height which is only half this maximum possible height. This observation thus verifies that all domains which are investigated in the current study are sufficiently long to accommodate, in principle, a fully-developped pattern without exceeding the limitations given in form of the angle of repose.

In the considered experimental studies, the shortest measured wavelength $\lambda/D \approx 105$ is clearly above the critical wavelength determined in the present work. In our simulations, the shortest box which allowed a single bedform to reach an asymptotic state with an asymmetric triangular shape was observed for a very similar box length $L_x/D = 102.4$. Nevertheless, the aspect ratio is still smaller than in the longer cases $H6$, $H7$ and $H12$ of \KU{}, which might be an indication that the patterns did not yet reach their final state and would further evolve if their growth was not hindered by the limited domain size.
Indeed, the majority of the experimental data points concentrates in a range of ${\lambda/D = 120-250}$ ($\lambda/H_f = 0.5-4.0$). In good agreement, all numerical simulations with $150 \leq L_x/D \leq 300$ develop patterns with a mean wavelength in the range ${150 \leq \lambda/D \leq 180}$ and a self-similar profile. These observations further support the suggestion of \KU{}, that the bedforms finally settle at a wavelength of approximately  ${150 \leq \lambda/D \leq 180}$. Similarly, \citet{Coleman_Nikora_2009,Coleman_Nikora_2011} report a lower bound for the preferred initial wavelength of $\lambda/D = 130$.
Eventually, the occurrence of patterns with a wavelength $\lambda/H_f < 3$ in numerical simulations and experiments which can be seen in figure~\ref{fig:overview_LxH_ov_LxD} excludes a possible scaling of the threshold with the mean fluid height which \KU{} could not exclude based on their limited data. This observation is further supported by the results of the analysis in the preceding section: cases with identical relative streamwise domain length $L_x/D$ but varying $L_x/H_f$ ratio attain similar final values for all studied geometrical parameters and spanwise-averaged bedform profiles.

Compared to the wavelengths observed in experiments, the initial wavelengths predicted by most linear stability analysis are smaller by at least one order of magnitude \citep{Langlois_Valance_2007,Ouriemi_2009}.
However, in some recent studies, the prediction of the most unstable wavelength could be improved taking additional physical effects into account. For instance, initial unstable pattern wavelengths similar to those determined in experiments were found after introducing an additional stabilizing effect in form of a phase-lag between the boundary shear stress and the particle flow rate into the linear stability analysis by several authors \citep{Charru_Hinch_2006,Charru_2006,Fourriere_al_2010}.
This phase shift, which is usually termed as characteristic saturation length $L_{sat}$, has been reported to be a function of the particle diameter, attaining values of the order of $10$ times the particle diameter for pure bedload transport \citep{Claudin_al_2011}.
For the special case of sand grains, \citet{Fourriere_al_2010} present saturation lengths between $7$ and $15$ grain diameters. In their subsequent stability analysis, these authors predict for sufficiently large fluid heights most amplified wavelengths in a range of $\lambda/L_{sat} = 15-20$, which is equivalent to a range $\lambda/D = 105-300$, when assuming that $L_{sat}  = 7-15$. This range of most amplified wavelengths overlaps with the values found in simulations and the mentioned experimental studies.

The findings of \citet{Colombini_Stocchino_2011}, on the other hand, differ from our observations and the results of \citet{Fourriere_al_2010}. For the range of Galileo numbers considered in the present study, the authors report only one unstable region that depends on the fluid height exclusively. In contrast, unstable wavelengths scaling with the particle diameter, which could be related to a ripple instability, appear in their model for $Ga \leq 14$ only. It should be however noted that the model of \citet{Colombini_Stocchino_2011} strongly depends on the roughness height and, consequently, on the relative submergence $H_f/D$, which is in their case at least one order of magnitude higher than in our numerical simulations. Therefore, a direct comparison of the predicted critical wavelengths with our results is not possible.

  \section{Conclusion}\label{sec:conclusion}
  %
%
%
In the current study, we have investigated the initial stage of subaqueous pattern formation in a turbulent open channel flow by means of direct numerical simulations with fully-resolved particles.
The main contribution is to address the question of scaling of the initial bedform wavelength with either the particle diameter $D$ or with the main fluid height $H_f$.
Both scaling relations have been proposed by different authors in the past decades, but up to the present day, there is no clear consensus about the correct scaling length.
In a recent paper, \citet{Kidan_Uhlmann_2017} have observed a lower bound for the most unstable wavelength $\lambda_{th}$ in a range  $75-100D$ (equivalent to $3-4H_f$, respectively). However, it was not possible in that study to further tackle the problem of the correct scaling, since the relative submergence $H_f/D$ was constant in all of their simulations.
In the present work, we have therefore performed and analysed two sets of simulations, in which we have varied the relative streamwise boxlengths $L_x/D$ and $L_x/H_f$ independently, which allowed us to investigate the influence of both length scales on the initial pattern wavelength exclusively. Consequently, the relative submergence has been varied in the range $H_f/D = 7.86 - 50.59$.

Our findings imply a scaling of the initial wavelength with the particle diameter, while it seems to be rather unaffected by variations of the mean fluid height. In particular, a lower bound for the most unstable wavelength has been found around a streamwise domain length of approximately $L_x/D = 80$. In cases with a shorter relative box length $L_x/D$, transverse pattern formation was effectively suppressed, indicating that the domain length is not sufficient to accommodate the minimal unstable wavelength.
However, the observed pattern with a wavelength close to the proposed threshold exhibit remarkable differences compared to patterns in sufficiently long domains (cf. for instance the cases with $L_x/D = \mathcal{O}(10^3)$ of \citet{Kidan_Uhlmann_2017}), i.e. in very marginal boxes it does not reach an asymptotic state during the observation time and shows a more symmetric spanwise-averaged profile than in the longer cases.

The predicted range of initial pattern wavelengths is in good agreement with values measured in laboratory channel and closed-conduits experiments, which concentrate predominantly in a range ${\lambda/D = 120-250}$. Furthermore, good agreement is also observed with wavelengths predicted as $\lambda/D = 105-300$ in the stability analysis of \citet{Fourriere_al_2010}. Note that in their study, the dependence of the most amplified wavelength on the particle diameter appears indirectly in form of a scaling with a saturation length $L_{sat}$, which in turn is a function of the particle diameter.

In contrast, the linear stability analysis of \citet{Colombini_Stocchino_2011} predicts only unstable wavelengths that scale with the fluid height for the range of Galileo numbers considered in the current study. It should be noted that since the relative submergence $H_f/D$, which is a crucial parameter in their study, is at least one order of magnitude larger than in our simulations, a direct comparison of predicted wavelengths from their stability analysis and the values observed in our simulations was not possible.
Generally, we could not observe such a scaling with the fluid height in our numerical simulations. For simulations with the same $L_x/D$ ratio, all relevant geometrical parameters attained comparable averaged values and profiles in the final phase irrespective of the varying $L_x/H_f$ ratio. In addition, initial wavelengths $\lambda/H_f < 3$ have been observed in experiments and simulations, excluding a scaling of the threshold which scales with the mean fluid height.

Another remarkable phenomenon has been observed in the cases with the highest relative submergence $H_f/D \approx 50$ in form of streamwise and spanwise sediment features, that are seen to either concur or interact with each other, allowing for the formation of three-dimensional patterns. The subsequent analysis of the single-sided amplitude spectra of the sediment bed height perturbation has shown that the evolution of the sediment bed is dominated by several streamwise and spanwise oriented waves of comparable amplitude. In the remaining simulations, streamwise aligned patterns appeared, if at all, mainly as a predecessor of the transverse patterns in the first few bulk time units of the simulations, but at clearly lower amplitude than their transverse oriented counterparts.
Since the focus of the current study was on the scaling of transverse patterns, it was out of the scope to investigate in detail the reasons for this different sediment bed evolution. However, it would be of high interest to determine the parameters that are responsible for the amplification of the spanwise waves. In this context, it should be investigated how initial subaqueous pattern formation changes when further increasing the Reynolds number.

  \section*{Acknowledgements} \label{sec:acknow}
  The current work was supported by the German Research Foundation (DFG) through grants
UH242/2-1 and UH242/12-1.
Part of the work was performed on the supercomputer ForHLR II at the Steinbuch Centre for Computing funded by the Ministry of Science, Research and the Arts Baden-W{\"u}rttemberg and by the Federal Ministry of Education and Research.
The remaining simulations have been carried out on SuperMUC at the Leibniz Supercomputing Centre at the Bavarian Academy of Science and Humanities. The computer resources, technical expertise and assistance provided by the staff at these computing centres are gratefully acknowledged.
We thank Michael Krayer for his support in defining the fluid-bed interface.

  \phantomsection
  \section*{Supplementary materials} \label{sec:suppmat}
  \addcontentsline{toc}{section}{Supplementary materials}
    Supplementary movies are available at
    \href{https://dx.doi.org/10.4121/uuid:7eb6a0be-ff83-4883-9d99-31daaa6a2863}{https://dx.doi.org/10.4121/uuid:7eb6a0be-ff83-4883-9d99-31daaa6a2863}.

  \appendix
  \section{Sensitivity to the choice of the Coulomb friction coefficient}\label{sec:append_A}
\begin{figure}[t]
    \centering
         \begin{minipage}{2ex}
           \rotatebox{90}
           {\small \hspace{5ex}$\sigma_h /D$}
         \end{minipage}
         \begin{minipage}{.44\linewidth}
           \raggedright{(\textit{a})}
           \includegraphics[width=\linewidth]
           {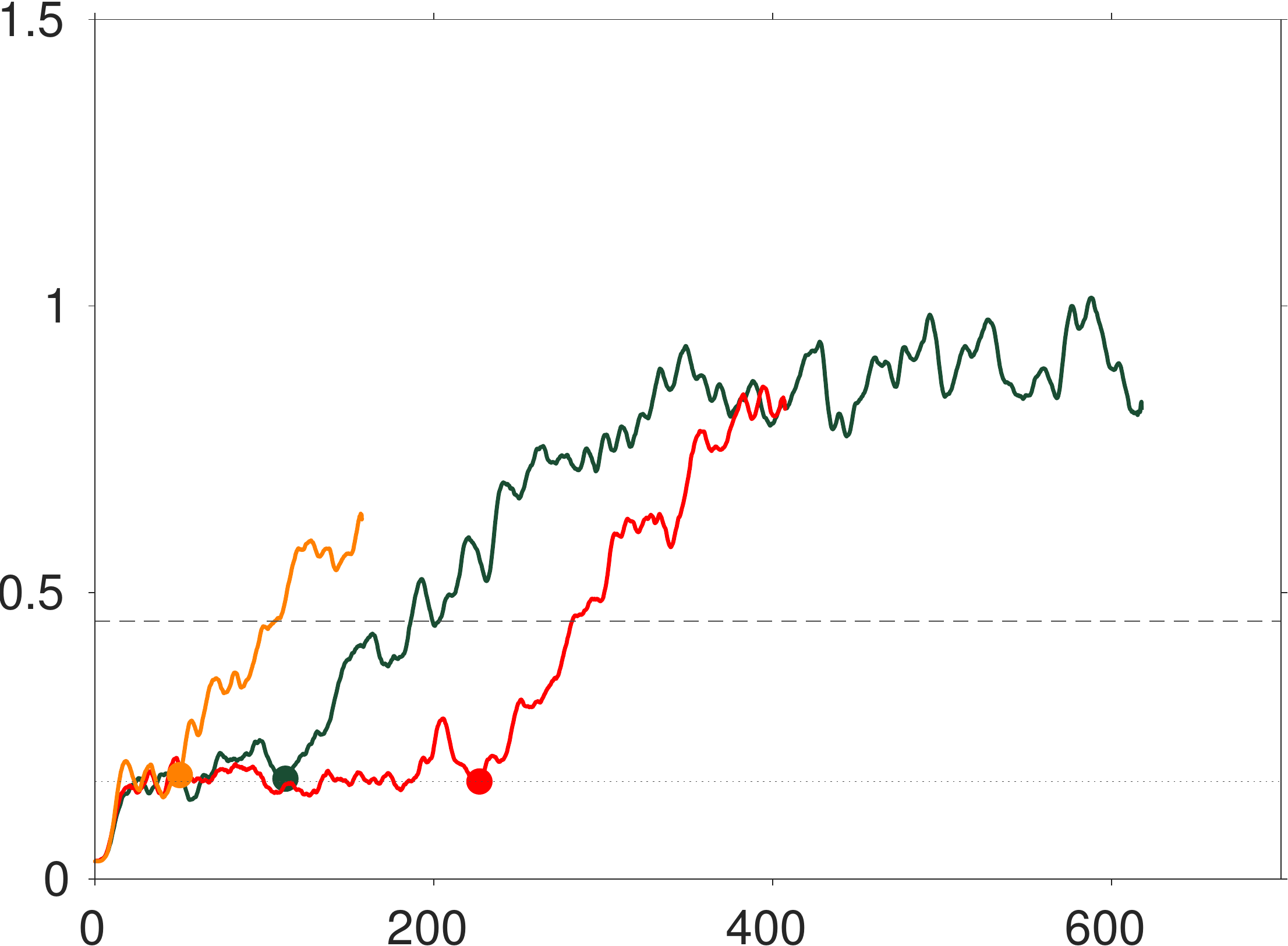}
           \centerline{\small $ t/(H_f/u_b) |_{\mu_c =0.5}$}
         \end{minipage}
         \begin{minipage}{.46\linewidth}
           \raggedright{(\textit{b})}
           \includegraphics[width=\linewidth]
           {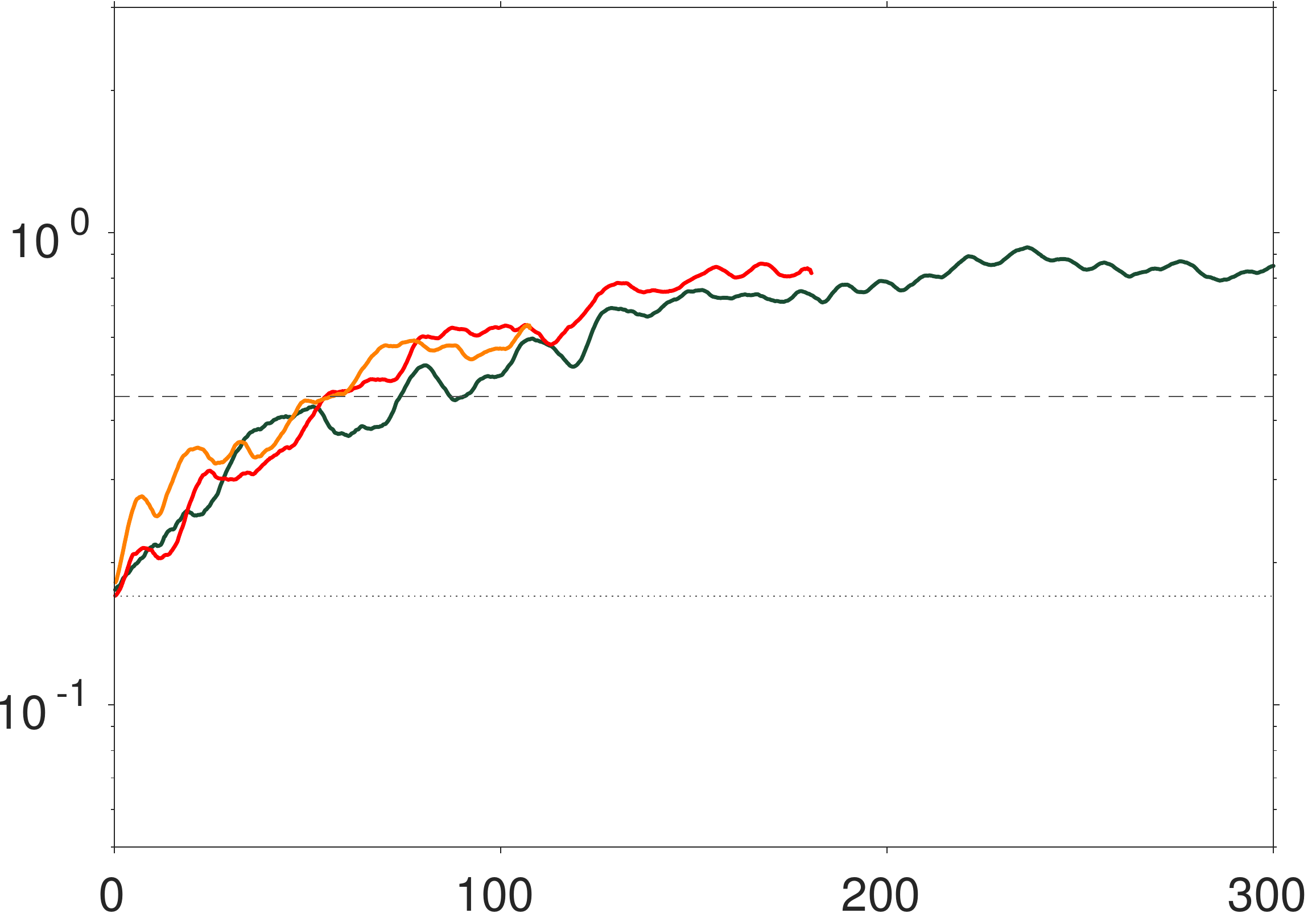}
           \centerline{\small $ (t-t^*)/(H_f/u_b) |_{\mu_c =0.5}$}
         \end{minipage}
  	\caption{
    (\textit{a})
    Time evolution of the root mean square of the bedform amplitude normalized by the particle diameter $\sigma_h /D$ for case \icaseD{} using different Coulomb friction coefficients:
    $\mu_c =0.5$ ({{\color{col_120} $\solidthick$}}),
    $\mu_c =0.4$ ({{\color{red} $\solidthick$}}),
    $\mu_c =0.4$ ({{\color{col_036} $\solidthick$}}, different initial flow field).
    (\textit{b})
    Same as in (\textit{a}), but in semi-logarithmic scale. The origin is shifted to the instant in time of the onset of the monotonic growth period $t^*$, which is indicated by filled circles in (\textit{a}).
    }
\label{fig:RMS1D_Dvar_comp_120_121_122}
\end{figure}

%
%
%
As mentioned in section~\ref{sec:numerics}, two of our performed simulations feature a slightly higher Coulomb friction coefficient $\mu_c=0.5$ compared to the remaining cases, where a value of $\mu_c=0.4$ was chosen. In the following, we show that the difference in this parameter has a minor effect on the eventually developed pattern and, in particular, that it does not affect the stability or instability of the sediment bed. To this end, we have recomputed case \icaseD{} keeping all parameters except for the Coulomb friction coefficient, which we reduced to the value $\mu_c=0.4$ as in the remaining simulations. Furthermore, we performed another simulation with identical parameters and $\mu_c=0.4$, but with a different initial flow field in order to check for the sensitivity to the initial data. 

In figure~\ref{fig:RMS1D_Dvar_comp_120_121_122}, we compare the time evolution of the root mean square sediment bed height fluctuation $\sigma_h$ for the three simulations.
It shows clearly that the bed becomes unstable for both values of $\mu_c$ and that, furthermore, the growth rate is very similar in all three simulations, although the instant in time of the onset of the monotonic growth regime differs. These observations indicate that the bedform evolution is not sensitive to a slight increase of the chosen $\mu_c$ in the observed range. Therefore, we will not further differ between the cases with $\mu_c=0.4$ and $\mu_c=0.5$ in the remainder of this study.

  \phantomsection
  \setlength{\bibsep}{.4ex}
  \addcontentsline{toc}{section}{References}
  \bibliography{literature.bib}

\end{document}